\documentclass[5p,times]{elsarticle}

\usepackage{amsmath,amssymb,amsfonts}
\usepackage{graphicx}
\usepackage{textcomp}
\usepackage[table]{xcolor}
\usepackage[ruled,linesnumbered]{algorithm2e}
\usepackage{algpseudocode}
\usepackage{commath}
\usepackage{multirow}
\usepackage{tabularx,booktabs}
\usepackage{multirow}
\usepackage{tikz}
\usepackage{caption}
\usepackage{subfigure}
\usepackage{placeins}
\usepackage{bm}
\usepackage{makecell}
\usepackage{hyperref}
\usepackage{balance}
\usepackage{amssymb}
\usepackage{pifont}

\usepackage{tcolorbox}

\DeclareMathOperator*{\argmin}{argmin}

\AtBeginDocument{%
  }
\makeatletter

\journal{Future Generation Computer Systems}

\newcommand{\MethodName}{\texttt{FedOptima}}

\newtheorem{theorem}{Theorem}

\newcolumntype{P}[1]{>{\centering\arraybackslash}p{#1}}

\RestyleAlgo{ruled}

\newif\ifshowchanges
\showchangesfalse
\newcommand{\di}[1]{\ifshowchanges{\color{blue} #1}\else#1\fi}

\tcbset{
    sharp corners,
    colback = white,
    before skip = 0.2cm,    
    after skip = 0.2cm      
}    

\newtcolorbox{boxH}{
    colback = gray!8, 
    colframe = black!60,
    boxrule = 0.5pt, 
    leftrule = 3pt,
    left=2pt,
    right=2pt,
    top=2pt,
    bottom=2pt
}

\begin{document}
\sloppy

\begin{frontmatter}

\title{
\texttt{FedOptima}: Optimizing resource utilization in federated learning
}

\author[sta]{Zihan Zhang\corref{cor1}}
\ead{zz66@st-andrews.ac.uk}
\author[rmi]{Leon Wong}
\ead{leon.wong@rakuten.com}
\author[sta]{Blesson Varghese}
\ead{blesson@st-andrews.ac.uk}

\cortext[cor1]{Corresponding author}

\affiliation[sta]{organization={School of Computer Science, University of St Andrews},
            addressline={Jack Cole Building}, 
            city={St Andrews},
            postcode={KY16 9SX}, 
            state={Fife},
            country={Scotland, United Kingdom}}

\affiliation[rmi]{organization={Autonomous Networking Research \& Innovation Department, Rakuten Mobile, Inc.},
            addressline={Rakuten Crimson House, 1-14-1 Tamagawa, Setagaya-ku}, 
            city={Tokyo},
            postcode={158-0094}, 
            state={Tokyo},
            country={Japan}}

\begin{abstract}
Federated learning (FL) systems facilitate distributed machine learning across a server and multiple devices. However, FL systems have low resource utilization on servers and devices, limiting their practical use in the real world. This inefficiency primarily arises from two types of idle time: (i) task dependency between the server and devices, and (ii) stragglers among heterogeneous devices. This paper introduces \MethodName, a resource-optimized FL system designed to simultaneously minimize both types of idle time; existing systems do not eliminate or reduce both at the same time.
\MethodName\ offloads the training of certain layers of a neural network from a device to a server using three innovations. 
First, devices operate independently of each other using asynchronous aggregation to eliminate straggler effects, and independently of the server by utilizing auxiliary networks to minimize idle time caused by task dependency. Second, the server performs centralized training using a task scheduler that ensures balanced contributions from all devices, improving model accuracy. Third, an efficient memory management mechanism on the server increases the scalability of the number of participating devices.
Extensive experiments are conducted on multiple lab-based testbeds, evaluated on image classification and sentiment analysis tasks with CNNs and Transformers. Compared to four state-of-the-art offloading-based and asynchronous FL baselines, \MethodName\ (i) achieves higher or comparable accuracy, (ii) accelerates training by 1.9$\times$ to 21.8$\times$, (iii) reduces server and device idle time by up to 93.9\% and 81.8\%, respectively, and (iv) increases throughput by 1.1$\times$ to 2.0$\times$.
\end{abstract}

\end{frontmatter}

\section{Introduction}
\label{sec:intro}
Federated learning (FL)~\cite{fedavg,DBLP:journals/corr/KonecnyMR15,DBLP:journals/corr/KonecnyMRR16} offers distributed training across user devices as an alternative to traditional centralized machine training. In FL, a server and multiple devices iteratively conduct local training and aggregation. Devices train a deep neural network (DNN) on their data and send model parameters to the server. The server aggregates these into a global model, which is then distributed to the devices for the next round. Thus, FL utilizes insight from user data via local models to train a global model without sharing the original data.

Sub-optimal resource utilization is a critical problem in FL that results in \textbf{two types of idle time} on the server and devices (see Section~\ref{subsec:problem}). The \textit{first is due to task dependency between server and devices}~- the server is idle for considerable periods when aggregating local models from devices as it waits for on-device training to complete, which is usually time-consuming. The \textit{second is due to hardware heterogeneity}~-faster devices are idle while waiting for slower devices (stragglers) that require more time to train.

Two categories of methods are considered in the existing literature for reducing idle time. Firstly, offloading-based FL (OFL) methods~\cite{splitfed,pipar,cse-fsl,fedgkt,10631047} improve server utilization by splitting the neural network and offloading the training of certain layers from the device to the server. Secondly, asynchronous FL (AFL) methods~\cite{fedasync,fedbuff,10631034} reduce the impact of stragglers by aggregating local models whenever the server receives them so that devices work independently of each other.

However, the above categories reduce idle time either on the server or devices, but not simultaneously (see Section~\ref{subsec:problem}). Simply combining OFL and AFL methods is inadequate as the combination inherits the limitations of both. As such, it is not scalable, and furthermore it introduces large communication overheads and imbalances the contributions of individual devices. These increase idle time and reduce model accuracy (Section~\ref{subsec:challenge}). To address these challenges, this paper proposes a resource utilization-optimized FL system, \MethodName. 

\textbf{Our solution}: \di{\MethodName\ proposes a training mechanism to simultaneously address Type~I and Type~II idle time (Section~\ref{subsec:challenge}). In particular, \MethodName\ combines (i) the asynchronous aggregation of models on the devices to mitigate stragglers, and (ii) gradient-free model offloading and centralized training on the server to mitigate server--device task dependency, where a straightforward combination of OFL and AFL is inadequate.} Each device trains a part of the DNN consisting of initial layers, which is made independent of the server by using an auxiliary network that computes a local loss. The on-device DNNs are aggregated asynchronously, ensuring that training on any device is independent of other devices. The server trains the rest of the DNN in a centralized manner to improve utilization and model accuracy. 

Evaluation on a range of models and datasets highlights that compared to the best results of four OFL and AFL baselines on multiple testbeds, \MethodName\ achieves higher or comparable accuracy, accelerates training by 1.9$\times$ to 21.8$\times$, reduces server and device idle time by up to 93.9\% and 81.8\%, respectively, and increases throughput by 1.1$\times$ to 2.0$\times$.
Our \textbf{research contributions} are:

1) An \textbf{idle time analysis} to precisely identify bottlenecks in FL, thereby informing targeted optimizations.

2) \di{A \textbf{gradient-free offloading mechanism} enabled by an auxiliary network, which eliminates the need to transfer gradients from the server-to-device and makes device-side training independent of the server while supporting asynchronous aggregation.}

3) \di{An \textbf{online task scheduling mechanism} based on server-side activation consumption counters to prevent fast devices from dominating centralized training and to balance device contributions under heterogeneity settings and non-IID data.}

4) \di{A \textbf{memory-bounded activation flow control mechanism} with a global buffering cap to decouple server-side activation buffering memory from the number of devices and improve scalability.}

The rest of this paper is organized as follows. Section~\ref{sec:motivation} discusses two types of idle time in FL and the challenges of addressing them simultaneously.
Section~\ref{sec:design} presents the system design of \MethodName. 
Section~\ref{sec:results} highlights the effectiveness of \MethodName\ compared to baselines. Related work is discussed in Section~\ref{sec:rw}, and conclusions are drawn in Section~\ref{sec:conclusion}.

\begin{table}[tp]
\di{
\centering
\caption{Summary of notations.}
\label{tab:notation}
\begin{tabular}{ll}
\hline
\textbf{Symbol} & \textbf{Meaning} \\
\hline
$K$ & Number of participating devices. \\
$k$ & Device index, $k \in [K]$. \\
$M$ & The original DNN to be trained. \\
$M_s$ & Server-side neural networks. \\
$M_d$ & Device-side neural networks. \\
$M_{dk}$ & Device-side network on Device $k$. \\
$\tilde{M}_{dk}$ & Auxiliary network on Device $k$. \\
$\theta_{dk}$ & Parameters of $M_{dk}$. \\
$\tilde{\theta}_{dk}$ & Parameters of $\tilde{M}_{dk}$. \\
$\theta_d$ & Global device-side model parameters. \\
$\tilde{\theta}_d$ & Global auxiliary model parameters. \\
$\theta_s$ & Parameters of the server-side model $M_s$. \\
$D_k$ & Local dataset on device $k$. \\
$\zeta$ & A mini-batch sampled from $D_k$. \\
$\xi$ & Activations produced by $M_{dk}$. \\
$\mathcal{A}$ & Activation set observed on the server. \\
$f_d(\cdot)$ & Device-side loss function. \\
$f_s(\cdot)$ & Server-side loss function. \\
$\mathcal{F}_d(\cdot)$ & Device-side objective. \\
$\mathcal{F}_s(\cdot)$ & Server-side objective. \\
$E_{d_{k}}$ & Maximum number of training rounds on Device $k$. \\
$H$ & Number of local iterations per round. \\
$t_k$ & Local model version index on Device $k$. \\
$t$ & Global model version index. \\
$\gamma_d$ & Device-side learning rate. \\
$\gamma_s$ & Server-side learning rate. \\
$D$ & Maximum delay in aggregation. \\
$\alpha$ & Aggregation weight. \\
$b_k$ & Network bandwidth of Device $k$. \\
$o_k$ & FLOP/s of device $k$. \\
$O_l, S^l$ & FLOPs of layer $l$ in $M$. \\
$S^l$ & Output size of layer $l$ in $M$. \\
$t_k^{train}$ & Training time on Device $k$. \\
$t_k^{transfer}$ & Activation transfer time on Device $k$. \\
$Q^{k}_{act}$ & Activation queue on server for Device $k$. \\
$\omega$ & Maximum queue length of $Q^{k}_{act}$. \\
$c_k$ & Activation counter for Device $k$. \\
$L$ & $L$-Smoothness assumption. \\
$\mu$ & $\mu$-weakly convex assumption. \\
$G$ & Upper bound of stochastic gradient norms. \\
$p^{t}(\xi)$ & Activation density at time $t$. \\
$p^{*}(\xi)$ & Converged activation density. \\
$q_t$ & Distance between $p^{t}(\xi)$ and $p^{*}(\xi)$. \\
$p$ & Probability of device dropout. \\
$T(p)$ & System throughput with $p$. \\
$R(p)$ & Retention ratio of system throughput with $p$. \\
\hline
\end{tabular}
}
\end{table}

\section{Background and motivation}
\label{sec:motivation}
This section presents an idle time analysis, followed by a discussion of the challenges and our proposed solution to simultaneously reduce server and device idle times. \di{Table~\ref{tab:notation} summarises the main symbols used throughout the paper.}

\subsection{Idle time analysis}
\label{subsec:problem}

Assume $K$ devices participate in FL training with their local dataset $\{\mathcal{D}_k;k\in [K]\}$. The objective of FL~\cite{fedavg} is:
\begin{equation}
\label{eq:fl}
    \underset{\bm{\theta}}{\min} \mathcal{F}(\bm{\theta}) = \frac{1}{k} \sum_{k=1}^{K} \mathbb{E}_{\zeta_k \sim \mathcal{D}_k} f_{k} (\bm{\theta})
\end{equation}
where $\bm{\theta}$ denotes the parameters of the global model and $\{f_k(\bm{\theta});k\in [K]\}$ denotes the local loss functions.

FL training involves the following steps: Step 1: \textit{Initialization} - a global model is initialized on a central server; Step 2: \textit{Local Training} - the global model is then distributed to devices, where each device trains the model on its local data; Step 3: \textit{Model Upload} - after local training, each device uploads its model updates (not the data) to the server; Step 4: \textit{Aggregation} - the server aggregates updates to improve the global model; Step 5: \textit{Model Distribution} - The updated global model is sent to the devices for the next training round.

The above steps are repeated over multiple iterations to improve the accuracy of the model. However, the iterative and distributed nature of FL results in low resource utilization, which introduces idle time. Two types of idle time are discussed below. 

\begin{figure*}[tp]
        \centering
        \subfigure[Federated Learning]{
	    \includegraphics[width=0.35\linewidth]{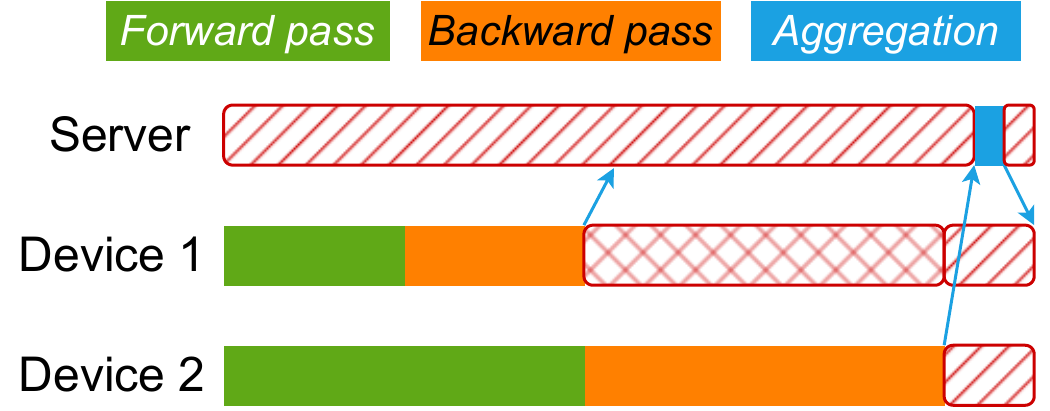}
	       \label{subfig:fl}
	    }
        \subfigure[Offloading-Based Federated Learning]{
	    \includegraphics[width=0.35\linewidth]{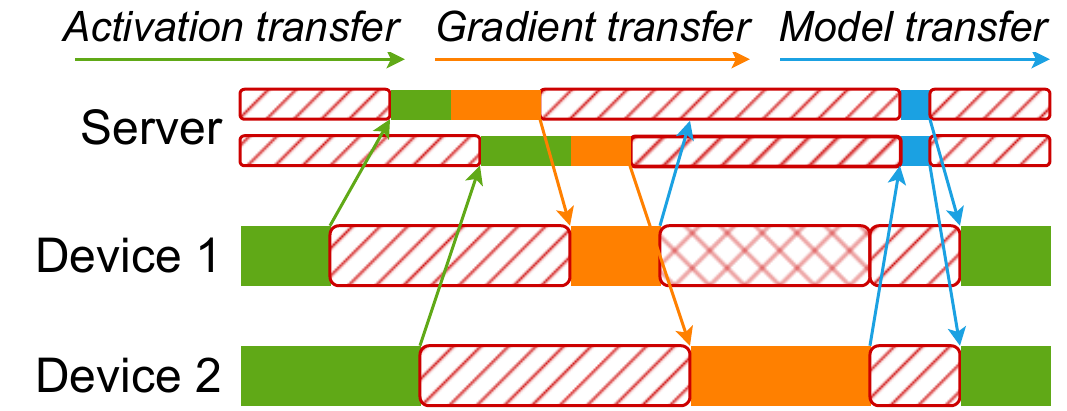}
	       \label{subfig:ofl}
	    }
        \subfigure[Asynchronous Federated Learning]{
	    \includegraphics[width=0.35\linewidth]{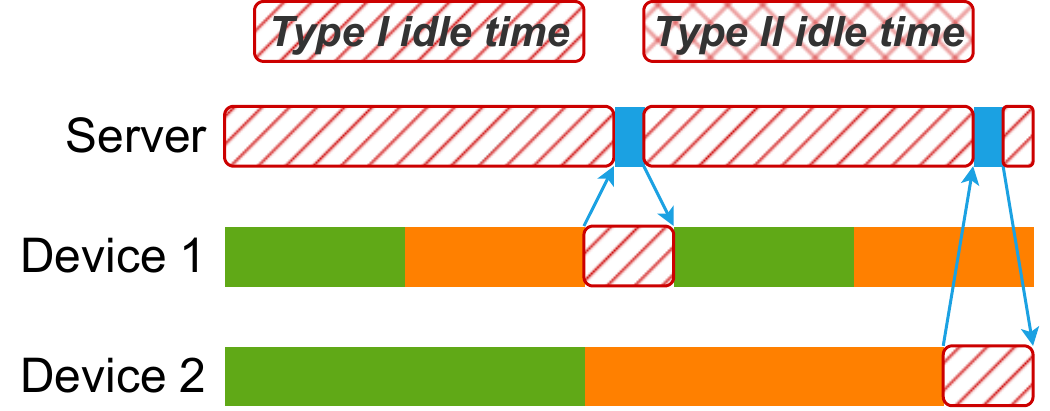}
	       \label{subfig:afl}
	    }
        \subfigure[\MethodName]{
	    \includegraphics[width=0.35\linewidth]{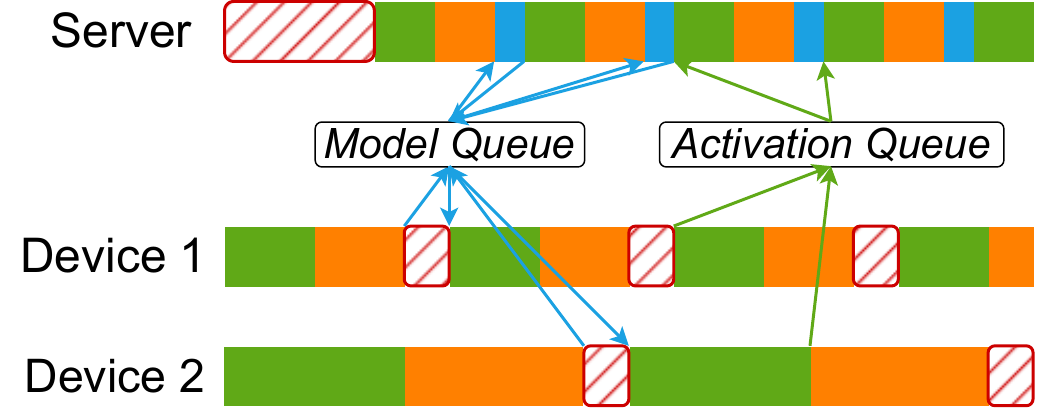}
	       \label{subfig:new-timeline}
	    }
	\caption{Training timeline for various federated learning (FL) methods with one server and two devices (Device 1 is assumed to be faster than Device 2). For simplicity, only one transfer of the model and activations per device is shown for \MethodName.}
        \label{fig:timeline}
\end{figure*}

\subsubsection{Idle time due to task dependency}

FL tasks train on the devices and aggregate on the server. A fundamental problem in FL is task dependency, which occurs since the server and device tasks depend on the results of each other. We define the idle time caused by exchanging the results of server and device tasks as \textbf{Type~I idle time}. In classic FL, device training and server aggregation tasks are interdependent. Aggregation requires the local models as input, while the training task needs the aggregated model as input. As shown in Figure~\ref{subfig:fl}, this dependency results in substantial idle periods on the server, as it must wait for all devices to complete training and send their models.

Offloading-based FL (OFL) methods, such as SplitFed~\cite{splitfed} and PiPar~\cite{pipar}, mitigate this problem by offloading training from the device to the server. In OFL, a DNN is partitioned, and the initial layers are trained on the device (referred to as device-side network), while the rest on the server (server-side network). 
The devices execute the forward pass of the device-side network and send intermediate results (or activations) to the server (Figure~\ref{subfig:ofl}). The server trains one server-side network corresponding to each device and sends the gradients back to the device to continue the backward pass and update the initial layers. This split architecture reduces Type~I idle time on the server. However, Type~I idle time increases on the device since it requires gradients from the server. Sending intermediate results between the devices and server increases the communication overhead. Although OFL improves server utilization, it does not address task dependency.

\subsubsection{Idle time due to stragglers}

Hardware heterogeneity among participating devices results in devices that slow down overall training. Faster devices wait for the stragglers to complete local training, giving rise to \textbf{Type~II idle time}. 

As shown in Figure~\ref{subfig:fl}, Device~1 waits for Device~2 to complete local training. Asynchronous FL (AFL) methods, such as FedAsync~\cite{fedasync} and FedBuff~\cite{fedbuff}, are proposed to address this. Aggregation in AFL requires one local model instead of models from all devices. The new model is the aggregation of the old global model and a local model. Devices send trained models to the server at their own pace, reducing Type~II idle time on the devices (Figure~\ref{subfig:afl}). However, faster devices exchange models with the server more frequently, increasing the communication overhead. 

Asynchronous aggregation also introduce the staleness problem. The local model may be stale when aggregated into the global model, because the latter may have already been updated by local models from other devices. This reduces model accuracy. AFL reduces the weight of the local model during aggregation as its staleness increases but the final accuracy is lower than classic FL.

\subsection{Challenges in addressing idle time}
\label{subsec:challenge}

Addressing Type~I and Type~II idle time simultaneously in FL is challenging. A straightforward solution might combine OFL and AFL methods. However, this combination inherits the limitations of the individual methods rather than mitigating them. This is demonstrated by developing an offloading-based asynchronous FL (OAFL) method that combines both. Specifically, the OAFL implementation is based on SplitFed~\cite{splitfed}, replacing the synchronous aggregation method with the asynchronous method used in FedAsync~\cite{fedasync}. 

A MobileNetV3-Large~\cite{mobilenet} model is trained on the Tiny ImageNet dataset~\cite{tinyimagenet} using OAFL on a testbed of 16 devices with heterogeneous computational resources (see the detailed setup in Section~\ref{subsec:setup}). Three challenges are encountered in training that motivate a new FL system. \MethodName\ is developed in response that combines the strengths of OFL and AFL and redesigns training to address the challenges. Below, we introduce these challenges and the solutions provided by \MethodName.

\subsubsection*{Challenge 1: Large communication overheads}
\label{ch:comm}


\begin{figure}[t]
    \centering\includegraphics[width=0.6\linewidth]{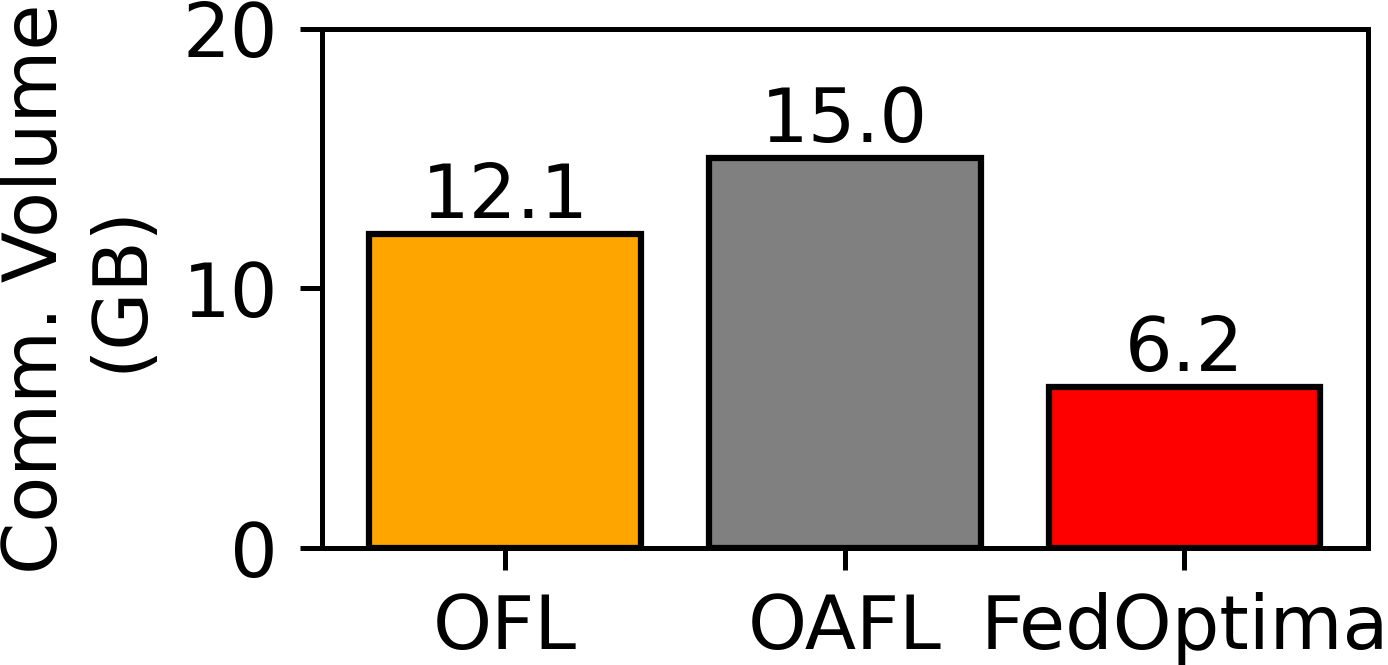}
    \caption{Communication volume per round.}
    \label{fig:comm-mot}
\end{figure}

\begin{figure}[t]
    \centering
    \includegraphics[width=0.6\linewidth]{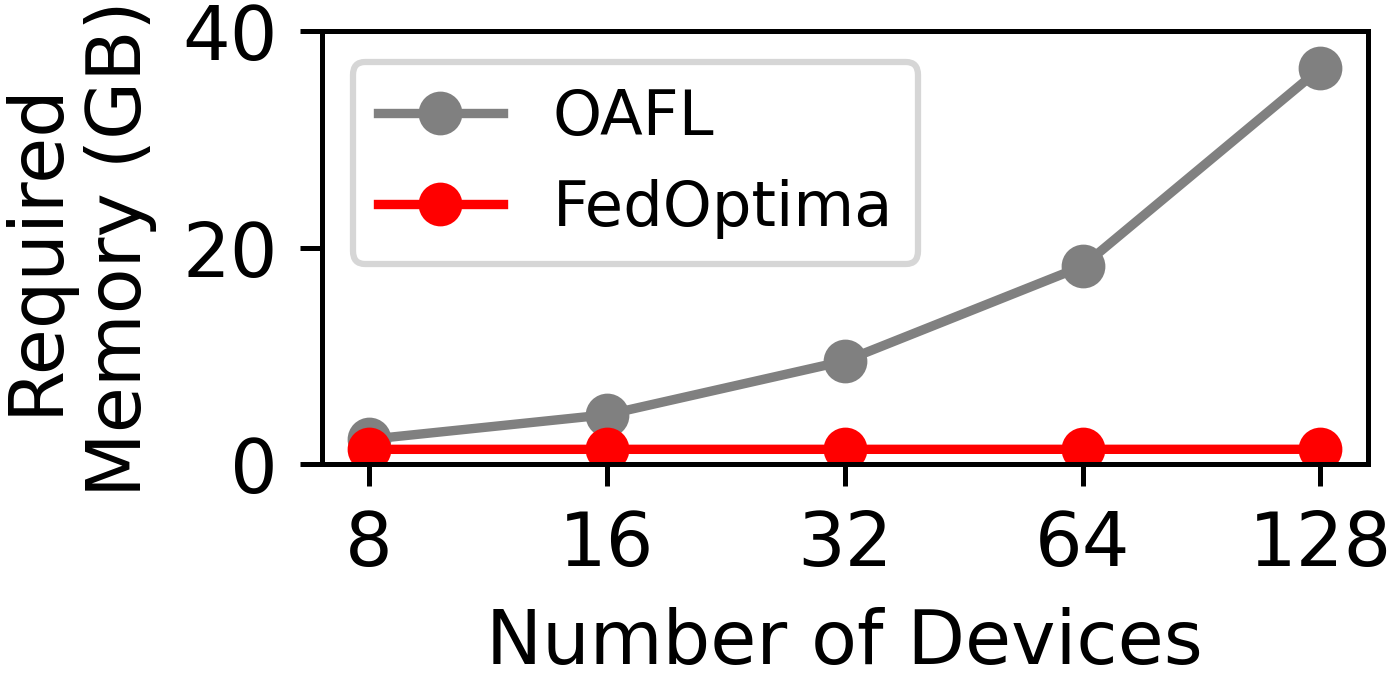}
    \caption{Memory required on server for varying no. of devices.}
    \label{fig:memory-mot}
\end{figure}

OFL methods, by design, require transferring activations and gradients between each device to the server in every training iteration. Due to synchronous aggregation, the faster devices pause exchanging data with the server when waiting for the stragglers. However, AFL methods allow faster devices to continue working without having to wait for the stragglers. Therefore, in OAFL all devices continuously exchange activations and gradients with the server without waiting for other devices, which increases the communication frequency.

The communication volume per round\footnote{A training round refers to training on $D$ data samples, where $D$ is the total size of datasets on all devices.} between devices and the server is measured during training. As shown in Figure~\ref{fig:comm-mot}, compared to OFL, OAFL has a 24\% increase in communication. Therefore, combining offloading-based and asynchronous methods makes communication a bottleneck rather than decreasing idle time. 

\di{\textbf{Our system:} Most communication in OFL and OAFL arises from transferring activations and gradients. \MethodName\ eliminates the need for a device to receive gradients from the server as each device trains its model independently with a local loss function (Section~\ref{subsec:device-design}). The activations are only transferred to the server upon request, rather than in every iteration, which further reduces the communication volume (Section~\ref{subsec:act-control}). Figure~\ref{fig:comm-mot} highlights that \MethodName\ has lower communication than both OFL and OAFL, which is consistent with our theoretical analysis. Specifically, \MethodName\ reduces communication of OAFL by 59\%, indicating an efficiency improvement in training.}

\subsubsection*{Challenge 2: Lack of scalability}
\label{ch:scale}

The available server memory limits the scalability of OFL, and consequently its combination with AFL. The maximum number of devices that can be trained depends on the number of server counterparts of the partitioned model that can be held in memory. As the number of devices increases, more server memory is required to accommodate their activations. 

The memory required $\mu$ by the server is as follows:
\begin{equation}
    \mu = (K+1) * \mu^{model} + K * \mu^{act}
\end{equation}
where $K$ denotes the number of devices, and $\mu^{model}$ and $\mu^{act}$ are sizes of model and activations, respectively. The first term of the equation indicates the server memory required for $K$ partitioned models and one global model. The second term represents the memory required by $K$ batches of activations.

Figure~\ref{fig:memory-mot} shows the server memory required by OAFL for varying number of devices. Memory linearly increases with the number of devices; OAFL is not scalable for a large number of devices. A server with 8~GB of memory can only work with at most 26 devices. Note that MobileNetV3-Large is a memory-efficient convolutional neural network (CNN), and other CNNs, such as VGG-16~\cite{vgg} or ResNet-101~\cite{resnet}, will further reduce the number of devices.

\textit{Our system:} In \MethodName, the server hosts a single model that receives activations from all devices (Section~\ref{subsec:server-design}). The server memory required is substantially lower than OAFL. The server has a predefined cap $\omega$ on the number of activations. Devices send activations only when the server allows (Section~\ref{subsec:act-control}). As a result, \MethodName\ operates on fixed memory budgets $\mu$ regardless of $K$, as shown in Equation~\ref{eq:memory-fedoptima} and Figure~\ref{fig:memory-mot}. Since the server model is trained in a centralized way (no aggregation required), \MethodName\ achieves comparable or higher accuracy than FL (Section~\ref{subsec:efficiency}).
\begin{equation}
\label{eq:memory-fedoptima}
    \mu = \mu^{model} + \omega * \mu^{act}
\end{equation}

\subsubsection*{Challenge 3: Disproportionate contribution}
\label{ch:unequal}

\begin{table}[tp]
    \centering
    \caption{The accuracy of MobileNet3-Large trained on Tiny ImageNet using homogeneous and heterogeneous devices.}
    \begin{tabular}{ccc}
        \Xhline{2\arrayrulewidth}
        \multirow{2}{*}{FL Method} & \multicolumn{2}{c}{Model Accuracy} \\
        \cline{2-3}
        & Homogeneous & Heterogeneous \\
        \Xhline{2\arrayrulewidth}
        OAFL & 24.025 & 20.02 \\
        \hline
        \MethodName & 39.75 & 40.05 \\
        \Xhline{2\arrayrulewidth}
    \end{tabular}
    \label{tab:homo-hetero}
\end{table}

When the server asynchronously aggregates local models in OAFL, faster devices that complete more training iterations contribute more to the global model than slower devices. 

The above affects the final model accuracy in two ways. Firstly, local models from slower devices may be stale as the global model may already be updated by local models from faster devices. In OAFL, staleness exists in both server and device models. Secondly, local models from slower devices are sent less frequently to the server, making a smaller contribution to the global model. Since device data in the real-world is non-independent and identically distributed (non-IID), the imbalanced contributions from different devices may lower the accuracy of the global model.

The impact on accuracy is shown in Table~\ref{tab:homo-hetero}. We use two testbeds - one comprising homogeneous devices and the other heterogeneous devices. The latter is simulated using homogeneous devices by changing the default GPU frequency (see Section~
\ref{subsec:setup}). Using OAFL, the accuracy of the model trained on the heterogeneous device platform is only 20.02\%, which is lower than the model accuracy on homogeneous devices (24.03\%).

\textit{Our system:} Since \MethodName\ trains only one global model it is never stale (Section~\ref{subsec:server-design}). The activations from different devices are selectively scheduled to the server, thereby balancing the contribution from different devices. Since the global model is trained centrally, the accuracy is higher than OAFL. Table~\ref{tab:homo-hetero} shows that models trained using \MethodName\ have comparable accuracy on homogeneous (39.75\%) and heterogeneous (40.05\%) devices, which is higher than the accuracy achieved by OAFL, suggesting that the model accuracy of \MethodName\ is not impacted by the disproportionate contribution of heterogeneous devices.

\subsubsection*{\textbf{Addressing Idle Time in Our System}}

In \MethodName, each device operates independently and transfers activations to the server periodically. The server places the activations into a queue and uses them to train a single server model. The device models are asynchronously aggregated in sequence. \di{These mechanisms are synergistic: the auxiliary network removes the server-to-device gradient dependency (enabling gradient-free offloading), asynchronous aggregation eliminates idle time on faster devices due to stragglers, and the server-side centralized training with scheduling and flow control ensures balanced contributions and bounded resource usage. This makes it possible to address Type~I and Type~II idle time simultaneously, which cannot be achieved by combining existing OFL and AFL methods. Finally, as shown in Figure~\ref{subfig:new-timeline}, \MethodName\ allows each device to work continuously without waiting for the server or other devices.}

\section{\texttt{F\MakeLowercase{ed}O\MakeLowercase{ptima}}}
\label{sec:design}
This section presents our proposed system \MethodName\ and how it reduces both types of idle time simultaneously.

\subsection{Overview}

\begin{figure*}[htp]
        \subfigure[Initialization]{
	    \includegraphics[width=0.45\linewidth]{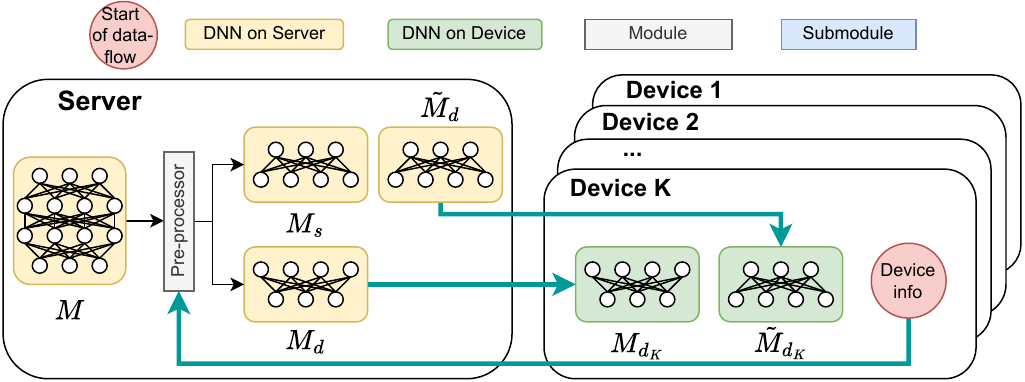}
	       \label{subfig:init}
	    }
        \subfigure[Training]{
	    \includegraphics[width=0.5\linewidth]{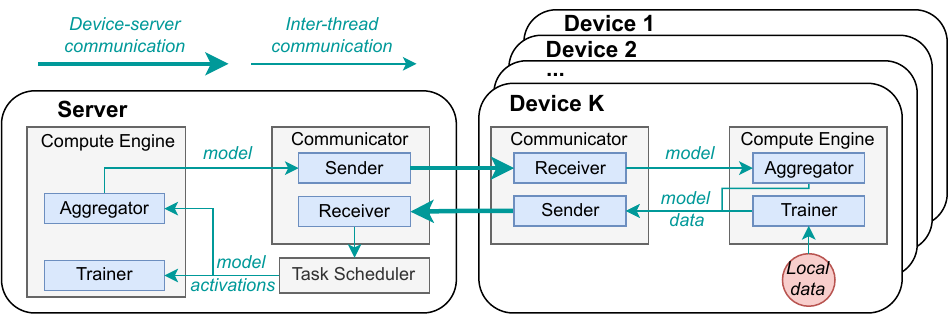}
	       \label{subfig:training}
	    }
	\caption{Overview of \MethodName.}
        \label{fig:overview}
\end{figure*}

\MethodName\ is a framework designed to improve FL training efficiency by reducing both Type~I and Type~II idle times. It consists of a server and $K$ devices, and it supports any sequential DNN, including convolutional neural networks (CNNs) and transformers. Figure~\ref{fig:overview} illustrates \MethodName\ with an initialization stage and a training stage.

During \textit{initialization} (Figure~\ref{subfig:init}), devices report system information (e.g., network bandwidth and floating-point operations per second) to the server. The server initializes the DNN $M$. A pre-processor module (Section~\ref{subsec:preprocessor}) uses the device information to partition $M$ into $M_d$ (the initial layers executed on devices) and $M_s$ (the remaining layers offloaded to the server). The pre-processor also generates an auxiliary network $\tilde{M}_d$ attached to the last layer of $M_d$. The pair sent to device $k$ is denoted by $(M_{d_k}, \tilde{M}_{d_k})$, referred to as the device-side network and auxiliary network, respectively, while $M_s$ is the server-side network. The auxiliary network $\tilde{M}_{d_k}$ is used for local training on devices. The server maintains the parameters of $M_d$ and $\tilde{M}_d$, denoted by $\bm{\theta}_d$ and $\bm{\tilde{\theta}}_d$, respectively.

\di{\textit{Training} in \MethodName\ is shown in Figure~\ref{subfig:training}. On device $k$, \textbf{device-side \MethodName} includes a Compute Engine (i.e., Trainer and Aggregator) and a Communicator (Section~\ref{subsec:device-design}). In repeated iterations, the Trainer processes a mini-batch of local data and outputs activations from the last layer of $M_{d_k}$. These activations are uploaded to the server to train $M_s$. Meanwhile, the Trainer uses $\tilde{M}_{d_k}$ and the local loss function $f_d$ to compute gradients locally, instead of receiving gradients from the server. This halves the communication volume and mitigates \textit{Challenge~1}. Here, we use the term ``model'' to refer to the parameters of a neural network (e.g., the parameters of $M_{d_k}$ are denoted by $\bm{\theta}_{d_k}$). After a predetermined number of iterations, the Aggregator forwards local models $\bm{\theta}_{d_k}$ to the server. The server aggregates them to obtain a global device-side model $\bm{\theta}_d$ (i.e., the parameters of $M_d$) and sends it back to devices to refresh local parameters. The main symbols are summarized in Table~\ref{tab:notation}.}

\di{\textbf{Server-side \MethodName} consists of a Communicator, a Task Scheduler, and a Compute Engine (Section~\ref{subsec:server-design}). The server receives activations from devices and uses them to train the server-side network $M_s$, which is analogous to centralized training but operates on activations instead of raw data. The server-side parameters $\bm{\theta}_s$ (i.e., the parameters of $M_s$) are iteratively updated by minimizing the server-side loss function $f_s$. To mitigate disproportionate contribution (\textit{Challenge~3}), the Task Scheduler selects activations for each iteration. \MethodName\ maintains a single server-side network and trains it using the scheduler-selected activations, mitigating \textit{Challenge~2}. The Task Scheduler also dispatches local models $\{\bm{\theta}_{d_k};k\in [K]\}$ to the Aggregator, which asynchronously aggregates them into the global device-side model $\bm{\theta}_d$ and sends updates back to devices. Due to asynchronous aggregation, staleness impacts the device-side model.}

\di{The overall objective of \MethodName\ consists of a device-side objective and a server-side objective, formulated in Equation~\ref{eq:fedoptima-device} and Equation~\ref{eq:fedoptima-server}, respectively. The local datasets on the $K$ devices are denoted by $\{\mathcal{D}_k; k\in [K]\}$, and the activation set on the server is denoted by $\mathcal{A}$. We define the device-side objective $\mathcal{F}_d$ as the expected device-side loss $f_d$ over local data samples $\zeta_k\sim\mathcal{D}_k$. Similarly, we define the server-side objective $\mathcal{F}_s$ as the expected server-side loss $f_s$ over activation samples $\xi\sim\mathcal{A}$.}

\begin{equation}
\label{eq:fedoptima-device}\underset{\bm{\theta}_d,\bm{\tilde{\theta}}_d}{\min} \mathcal{F}_d(\bm{\theta}_d,\bm{\tilde{\theta}}_d) = \frac{1}{K} \sum_{k=1}^{K} \mathbb{E}_{\zeta_k \sim \mathcal{D}_k} f_d (\bm{\theta}_d,\bm{\tilde{\theta}}_d; \zeta_k)
\end{equation}
\begin{equation}
\label{eq:fedoptima-server}
    \underset{\bm{\theta}_s}{\min} \mathcal{F}_s(\bm{\theta}_s) = \mathbb{E}_{\xi \sim \mathcal{A}} f_s(\bm{\theta}_s; \xi)
\end{equation}


\subsection{Initialization stage}
\label{subsec:preprocessor}

In the initialization stage, the devices share information, namely network bandwidth $b_k$ and CPU/GPU floating-point operations per second $o_k$, with the server. The \textbf{Pre-processor} uses the device information to (1) split the DNN $M$ into $M_d$ and $M_s$, and (2) generate an auxiliary network.

\subsubsection{Splitting the DNN}

When splitting $M$, the Pre-processor selects sequential neural network layer connections as split points and does not cut off the branch network. In selecting the split point, the trade-off between device training time $t^{train}$ and activation transfer time $t^{transfer}$ are considered. $M$ is profiled by the Pre-processor to obtain the floating operations $O_l$ and the output sizes of each layer $\{S_l, l\in [L]\}$, where $L$ is the layer number of $M$. Then $t^{train}$ and $t^{transfer}$ with split point $l$ for device $k$ are estimated by Equation~\ref{eq:time} and Equation~\ref{eq:transfer_time}, respectively.
\begin{equation}
\label{eq:time}
    t^{train}_k(l) = \sum_{i=1}^{l} O_i/o_k
\end{equation}
\begin{equation}
\label{eq:transfer_time}
    t^{transfer}_k(l) = S_l/b_k
\end{equation}

The selection of the split point $l$ is formulated as follows:
\begin{equation}
    l = \argmin_{l=1}^L \max_{k=1}^K \max\{t^{train}_k(l), t^{transfer}_k(l)\}
\end{equation}

\subsubsection{Designing the auxiliary network}
\label{subsubsuc:aux}

The auxiliary network $\tilde{M}_{d_k}$ serves as an extension to the local network $M_{d_k}$, specifically designed to adapt the output of the network for local objectives $f_d$ to provide gradients for model updates. Compared to existing offloading FL methods, using an auxiliary network allows the device to update the model without receiving gradients from the server, thus significantly reducing communication traffic. $\tilde{M}_{d_k}$ usually consists of fewer layers than $M_s$. The default option of $\tilde{M}_{d_k}$ is one layer of the same type as the last layer of $M_{d_k}$, followed by a dense layer as the classifier. We use the default option in experiments. The layer number of $\tilde{M}_{d_k}$ is a tunable hyperparameter.

The overhead of the initialization stage is under 5 seconds.

\subsection{Training stage}

The $K$ devices and a server collaborate to train a DNN in \MethodName, which are presented in this section.

\subsubsection{Device-side \MethodName}
\label{subsec:device-design}

Each device has two modules that run in parallel: the Compute Engine and the Communicator.

The \textbf{Communicator} uses a Sender and Receiver in parallel that maintains a sending queue for outgoing messages and a receiving queue for incoming messages. The Sender queues activations and local models for dispatch. Incoming messages queued in the Receiver are retrieved by the Compute Engine. 

\begin{algorithm}[t]
    \caption{Compute Engine on device $k$.}
	\label{alg:device-compute}
	\KwIn{Device-side network $M_{d_k}$ with parameters $\bm{\theta}_{d_k}$; auxiliary network $\tilde{M}_{d_k}$ with parameters $\bm{\tilde{\theta}}_{d_k}$; local dataset $\mathcal{D}_k$; maximum local round number $E_{d_k}$; iteration no. per round $H$; device learning rate $\gamma_d$}
	\KwOut{Device-side model $\bm{\theta}_{d_k}$}
	\BlankLine
        Let local model version $t_k \gets 0$
        
        \For{$e \gets 0$ to $E_{d_k}$}{
            \tcc{Local training iterations.}
            \For{$h \gets 0$ \KwTo $H$}{
                Get a mini-batch of data $\zeta_k^h$ from $\mathcal{D}_k$

                Execute forward pass of $M_{d_k}$ on $\zeta_k^h$

                Get activations $\xi_k^h$ and put into the Sender
                
                Execute forward pass of $\tilde{M}_{d_k}$ on $\xi_k^h$

                Get local loss $f_d (\bm{\theta}_{d_k},\bm{\tilde{\theta}}_{d_k}; \zeta_k^h)$

                Execute backward pass of $\tilde{M}_{d_k}$ and $M_{d_k}$

                Let $\bm{\theta}_{d_k} \gets \bm{\theta}_{d_k} - \gamma_d \partial f_d / \partial \bm{\theta}_{d_k}$

                Let $\bm{\tilde{\theta}}_{d_k} \gets \bm{\tilde{\theta}}_{d_k} - \gamma_d \partial f_d / \partial \bm{\tilde{\theta}}_{d_k}$
            }
            \tcc{Aggregation.}
            Put $\{\bm{\theta}_{d_k}, \bm{\tilde{\theta}}_{d_k}, t_k\}$ into the Sender

            Get $\{\bm{\theta}_d, \bm{\tilde{\theta}}_d, t\}$ from the Receiver

            Let $\{\bm{\theta}_{d_k}, \bm{\tilde{\theta}}_{d_k}, t_k\} \gets \{\bm{\theta}_d, \bm{\tilde{\theta}}_d, t\}$

            \uIf{Receive STOP}{
                \textit{break}
            }
        }
        \textbf{Return} $\bm{\theta}_{d_k}$
\end{algorithm}

The \textbf{Compute Engine} has two submodules: a Trainer for iteratively training the local device-side model and an Aggregator for periodically aggregating the local model. As shown in Algorithm~\ref{alg:device-compute}, the device-side network is trained in multiple rounds ($E_{d_k}$ rounds if no stop signal is received), each consisting of $H$ iterations (Lines~2-18). The device-side model version $t_k$ is initialized as 0 at the start (Line~1). Within each iteration (Lines~3-12), the Trainer processes a mini-batch of data through $M_{d_k}$ (Line~4). After the forward pass (Line~5), the activations are placed with the Sender for server-side training (Line~6). Meanwhile, the activations are used to compute the local loss through $\tilde{M}_{d_k}$ (Lines~7-8). Backpropagation follows (Line~9), leveraging the local loss to calculate gradients that update the model parameters (Lines~10-11) required for the next iteration. 
The Aggregator works at the end of each round. The device-side model and the auxiliary model, along with their version, are put into the Sender (Line~13). After the aggregation on the server completes, the new models are received from the server (Line~14) and replace the old ones (Line~15).
Training completes when a stop signal is received (Line~16) or after $E_{d_k}$ rounds are completed. 

\subsubsection{Server-side \MethodName}
\label{subsec:server-design}

The server trains the server-side model using activations from devices and aggregates device-side models. The server has three modules: the Communicator, the Task Scheduler, and the Compute Engine.

All devices are concurrently connected to the \textbf{Communicator}. The Sender and Receiver maintain a sending queue and a receiving queue, respectively, accessed by the Task Scheduler.



The \textbf{Task Scheduler} handles two message types: local models and activations. The $put()$ and $get()$ algorithms re-order received messages from the Communicator before dispatching them to the Compute Engine when requested. 

\begin{algorithm}[t]
    \caption{$put()$ algorithm of the Task Scheduler.}
	\label{alg:put}
	\KwIn{Received message $m$; model queue $Q^{model}$; activation queues $\{Q^{act}_k; k\in [K]\}$
	    }
	\BlankLine
	\uIf{$m$.type == ``model''}{
            Let $\{\bm{\theta}_{d_k}, \bm{\tilde{\theta}}_{d_k}, t_k\} \gets m.content$
            
            Put $\{\bm{\theta}_{d_k}, \bm{\tilde{\theta}}_{d_k}, t_k\}$ to $Q^{model}$
        }
        \Else{
           Let $\xi \gets m.content$

           Let $k \gets m.origin$
            
            Put $\xi$ to $Q^{act}_k$
        }
        \textbf{Return}
\end{algorithm}

The Task Scheduler maintains a model queue $Q^{model}$ to store models and $K$ activation queues $\{Q^{act}_k, k=1, \cdots, K\}$ to store activations received from each device. As shown in Algorithm~\ref{alg:put}, the Task Scheduler recognizes whether a message from the Communicator contains a combination of the local models (Line~1) or a batch of activations (Line~4). The local models and activation batches are then put into the corresponding queues (Lines~2-3, 5-7).

\begin{algorithm}[t]
    \caption{$get()$ algorithm of the Task Scheduler.}
	\label{alg:get}
	\KwIn{Model queue $Q^{model}$; activation queues $\{Q^{act}_k; k\in [K]\}$; activation counter $\{c_k; k\in [K]\}$
	    }
	\KwOut{Activations $\xi$ or local models $\{\bm{\theta}_{d_k}, \bm{\tilde{\theta}}_{d_k}, t_{d_k}\}$}
	\BlankLine
	\uIf{$Q^{model}$ is not empty}{
            Get $\{\bm{\theta}_{d_k}, \bm{\tilde{\theta}}_{d_k}, t_k\}$ from $Q^{model}$
            
            \textbf{Return} $\{\bm{\theta}_{d_k}, \bm{\tilde{\theta}}_{d_k}, t_k\}$
        }
        \Else{
           Let $k \gets \argmin_k c_k$

           Get $\xi$ from $Q^{act}_k$

           Let $c_k \gets c_k+1$

           \textbf{Return} $\xi$
        }
\end{algorithm}

When a request is received from the Compute Engine, the Task Scheduler gets a batch of activations for training or a combination of local models for aggregation. As shown in Algorithm~\ref{alg:get}, a higher priority is given for retrieving models (Line~1) to the Compute Engine over the activations. If there are no pending models in the queue, the Task Scheduler retrieves activations for the Trainer to balance training continuity and timely model aggregation (Line~4). The FIFO policy is used to obtain local model combinations from the queue (Lines~2-3). To address Challenge~3, a counter-based sorting algorithm is implemented to get activations (Lines~5-8). The counter $\{c_k; k\in [K]\}$ tallies the activations used for training from each device and is used to prioritize devices that contributed fewer activations. \di{$c_k$ is the number of activation batches from device $k$ that are consumed by the server for training. In each scheduling step, the Task Scheduler selects activations from the device with the smallest $c_k$. This online greedy strategy keeps the consumption counts $\{c_k; k\in [K]\}$ nearly balanced over time and prevents activations of fast devices from dominating server-side training.} 

\begin{algorithm}[t]
    \caption{Compute Engine on server.}
	\label{alg:server-compute}
	\KwIn{Device no. $K$; server-side network $M_s$ with parameters $\bm{\theta}_s$; maximum global round no. $E_s$; server learning rate $\gamma_s$; maximum delay $D$; aggregation weight $\alpha$}
	\KwOut{Server-side model $\bm{\theta}_s$}
	\BlankLine
        Let global model version $t \gets 0$
        
        \For{$e \gets 0$ \KwTo $E_s$}{
            \While{$m \gets$ Task Scheduler}{                
                \uIf{$m$.type == ``activation"}{
                    \tcc{Global training iterations.}
                    Let activations $\xi \gets m.content$

                    Execute forward pass of $M_s$

                    Get global loss  $f_s (\bm{\theta}_s; \xi)$

                    Execute backward pass of $M_s$

                    Get gradients $\nabla f_s (\bm{\theta}_s; \xi)$

                    $\bm{\theta}_s \gets \bm{\theta}_s - \gamma_s \nabla f_s (\bm{\theta}_s; \xi)$
                }
                \Else{
                    \tcc{Aggregation.}
                    Let $\{\bm{\theta}_{d_k}, \bm{\tilde{\theta}}_{d_k}, t_k\} \gets m.content$

                    \If{$t - t_k > D$}{
                        \textit{continue}
                    }

                    Let $\alpha \gets \frac{1}{t-t_k+1}$
                    
                    Let $\bm{\theta}_d \gets \alpha\bm{\theta}_{d_k} + (1-\alpha)\bm{\theta}_d$
                    
                    Let $\bm{\tilde{\theta}}_d \gets \alpha\bm{\tilde{\theta}}_{d_k} + (1-\alpha)\bm{\tilde{\theta}}_d$

                    Let $t \gets t + 1$

                    Put $\{\bm{\theta}_d, \bm{\tilde{\theta}}_d, t\}$ into the Sender

                    \uIf{$t \% K == 0$}{
                        \textit{break}
                    }
                }
            }
            \uIf{$\bm{\theta}_s$ has converged}{
                \textit{break}
            }
        }
        Send \textit{STOP} to all devices
        
        \textbf{Return} $\bm{\theta}_s$
\end{algorithm}

The \textbf{Compute Engine} requests inputs continuously from the Task Scheduler to carry out training and aggregation. As shown in Algorithm~\ref{alg:server-compute}, training consists of at most $E_s$ rounds. The global device-side model version is set to 0 at the beginning (Line~1). Within each training round (Lines~2-28), the Compute Engine requests a message repeatedly from the Task Scheduler (Line~3). When activations are received (Line~4), one training iteration of the server-side network $M_s$ is executed using stochastic gradient descent (Lines~5-10), analogous to centralized training but using activations instead of the original data. This addresses server-side model staleness. Upon receiving a local model combination (Line~11), the Compute Engine performs aggregation (Lines~12-19) using FedAsync~\cite{fedasync}. To reduce the impact of staleness, aggregation does not occur if the difference between the versions of the local model and the global model is higher than the predefined maximum delay $D$ (Lines~13-14). Otherwise, the aggregation weight $\alpha$ is updated based on the difference of the version (Line~16), and the global model, global auxiliary model and global model version are updated (Lines~17-19) and put into the Sender (Line~20). A global training round is completed after finishing $K$ aggregation (Lines~21-22). Then the next round commences. 
After the server-side model converges or $E_s$ rounds, the training completes (Lines~25-27) and a stop signal is sent to all devices (Line~28).

\subsection{Memory and device management}
\label{subsec:act-control}

The section presents the activation flow control mechanism that prevents server memory overflow and the device management during training.

\subsubsection{Activation flow control}

\begin{figure*}[tp]
	\centering
    \includegraphics[width=0.55\linewidth]{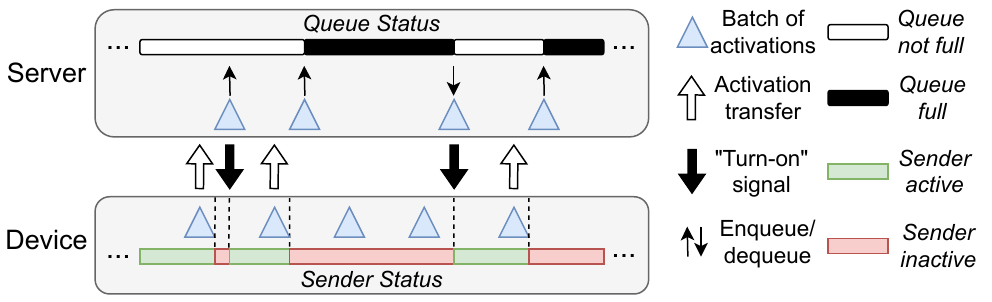}
	\caption{The activation flow control between a device and the server. A device only sends activations to the server when the Sender is active.}
        \label{fig:act-flow}
\end{figure*}

Efficient memory management on the server is required in \MethodName\ due to the continuous transfer of activations from devices.
Each device maintains a \textit{Sender Status} to control the transfer of activations to the server. The server manages activation queues for each device, denoted as $Q^{act}_k$ for device $k$. \di{To decouple server-side memory usage from the number of devices, we enforce a \textbf{global} activation buffering cap $\omega$ (an upper bound on the total number of activation batches buffered across all devices), i.e., $\sum_{k=1}^K |Q^{act}_k| \leq \omega$.} The activation flow control is as follows (see Figure~\ref{fig:act-flow}).

\textit{Device-side Flow Control}: After every local training iteration, each device evaluates its \textit{Sender Status} before sending activations. If the Sender is active, the device sends activations to the server and then deactivates the Sender until receiving a `turn-on' signal from the server; otherwise, the device proceeds to the next iteration without sending activations.

\textit{Server-side Flow Control}: The server adds the activations received to the queue of the corresponding device $Q^{act}_k$. When activations are enqueued to or dequeued from $Q^{act}_k$, the server checks whether the \di{global buffering cap $\omega$} is reached. If not, it dispatches a `turn-on' signal to the corresponding device, reactivating its Sender.

\di{$\omega$ is determined by the memory available on the server, and is independent of the number of devices. A large $\omega$ reduces backpressure (devices are allowed to upload activations more frequently), which improves server-side training continuity and utilization, but consumes more memory. A smaller $\omega$ reduces server memory usage, but may increase activation starvation at the server (i.e., fewer activations available for training at a given time), which can reduce throughput and increase time to achieve model convergence; however, since devices repeatedly iterate over their local data during training, activations corresponding to each data point can still be observed by the server over the course of training. For a large number of devices with changing network conditions, the system remains memory-safe due to the global cap; however, as more devices compete for a fixed $\omega$, the effective buffer shared per device decreases and the system throughput will be bound by server compute/network resources and the chosen $\omega$.}


\subsubsection{Devices joining and leaving}

Devices may not be always available in real-world FL. In synchronous FL, devices that participate are selected before a training round to reduce delays due to failed or dropped devices~\cite{fedavg,oort,refl}. Since local device models are aggregated asynchronously in asynchronous FL, devices that join or leave do not delay training~\cite{fedasync,fedbuff}.

However, the trade-off between system and statistical performance needs to be considered when selecting participating devices~\cite{oort,refl}. 
Faster devices are more likely to be selected for improving system performance, but the data on slower devices will be less represented in the global model. This results in requiring a larger number of training rounds to achieve a target model accuracy. A wider coverage of data is required for improving statistical performance but requires slower devices to be selected for training, which will inevitably increase the training time per round. 

In \MethodName, since devices work independently they join or leave without interrupting training. When a device leaves training, the other devices and server will continue training to maintain system performance. Activation flow control discussed above ensures that new devices that join do not contend for server resources. The Task Scheduler in \MethodName\ balances data across different devices, thereby improving statistical performance in FL. In short, \MethodName\ balances system performance and statistical performance.

\di{\subsection{Limitations and Deployment Considerations}} 
FL, AFL and OFL inherently enhance privacy by transferring models or activations instead of the original data. Established techniques, such as differential privacy~\cite{dp, pixeldp}, mitigate the potential risk of reconstructing data from models or activations. \di{Activation transfer may still pose privacy risks (e.g., reconstruction/inversion attacks on intermediate representations), and therefore \MethodName\ does not provide strong privacy guarantees by default.}

While privacy is not the main focus of this paper, we note that \MethodName\ is compatible with existing model-level protection mechanisms, such as differential privacy (DP)~\cite{dp}, and activation-level protection mechanisms, such as PixelDP~\cite{pixeldp}. In \MethodName, the activations transferred from devices to the server have the same structure and semantics as those in existing OFL systems. \di{Although \MethodName\ increases the amount of server-side computation for efficiency, it only exposes the same type of information to the server as in split learning/offloading-based FL: the original data still remain on devices and the server only observes activations generated at a chosen split point.}

Therefore, standard techniques like inserting a PixelDP noise layer before upload can be readily applied, and the resulting privacy--utility trade-offs would be comparable to prior work. \di{For sensitive deployments, \MethodName\ should be combined with such defenses and, where applicable, additional measures such as encryption/secure enclaves to meet the required privacy level.}

As our contribution focuses on system efficiency and scheduling, we do not further consider privacy-preserving techniques in this paper.

\section{Convergence analysis}
\label{sec:converge}
The section analyses the convergence of the device-side model $\bm{\theta}_d$ and the server-side model $\bm{\theta}_s$ in \MethodName. \di{The analysis is based on existing contributions in the area: the device-side model follows standard asynchronous FL aggregation~\cite{fedasync}, and the server-side model follows existing results on training subsequent layers from intermediate activations~\cite{decoupled}. Our goal is to clarify that existing theorems apply to \MethodName's training pipeline, thereby complementing the empirical convergence behavior observed in Section~\ref{sec:results}.}

The basic assumptions~\cite{fedasync, decoupled} are made as follows:

\textbf{Assumption 1} (Smoothness): The differential loss functions on devices $f_d$ and the server $f_s$ are L-smoothness. \di{We use $f$ to denote a generic loss function and is instantiated as either the device-side loss $f_d$ or the server-side loss $f_s$, i.e. $f \in \{f_d, f_k\}$. There exists $L>0$ such that for $\forall \bm{\theta}_1,\bm{\theta}_2$, we have}
\begin{equation}
    f(\bm{\theta}_2) - f(\bm{\theta}_1) \leq  \langle \nabla f(\bm{\theta}_1) , \bm{\theta}_2 - \bm{\theta}_1 \rangle + \frac{L}{2} \lVert \bm{\theta}_2 - \bm{\theta}_1 \rVert 
\end{equation}

\textbf{Assumption 2} (Weak convexity): The differential loss functions on devices $f_d$ and the server $f_s$ are $\mu$-weakly convex. There exists $\mu>0$ and a convex function $g$ with $g(\bm{\theta}) = f(\bm{\theta}) + \frac{\mu}{2} \lVert \bm{\theta} \rVert^2$ for \di{$f \in \{f_d, f_k\}$} such that
\begin{equation}
    g(\bm{\theta}_2) - g(\bm{\theta}_1) \geq  \langle \nabla g(\bm{\theta}_1) , \bm{\theta}_2 - \bm{\theta}_1 \rangle
\end{equation}

\textbf{Assumption 3} (Bounded gradients): There exists $G > 0$ such that for $\forall \bm{\theta}_d, \bm{\tilde{\theta}}_d, \bm{\theta}_s$, $k\in [K]$, $\forall \zeta \sim \mathcal{D}_k$ and $\forall \xi \sim \mathcal{A}$, we have
\begin{equation}
    \lVert \nabla f_d(\bm{\theta}_d, \bm{\tilde{\theta}}_d; \zeta) \rVert^2 \leq G \quad \text{and} \quad  \lVert \nabla f_s(\bm{\theta}_s; \xi) \rVert^2 \leq G
\end{equation}

\textbf{Assumption 4} (Robbins-Monro conditions): The server-side learning rate $\gamma_s$ satisfy
\begin{equation}
    \sum_t \gamma_s^t = \infty \quad \text{and} \quad \sum_t |\gamma_s^t|^2 < \infty, 
\end{equation}

\subsection{Device-side model convergence}

The device-side model is trained in a fully asynchronous manner, which is same as FedAsync~\cite{fedasync}. According to Assumption 1, 2 and 3 and Theory 5 presented in literature~\cite{fedasync}, we directly have

\begin{theorem}[Device Block Convergence]
For any small content $\forall \epsilon > 0$ the device-side model converges to a critical point:
\begin{equation}
\label{eq:device-converge}
    \begin{split}
        \min_{t=0}^{T-1} \mathbb{E}\lVert \nabla \mathcal{F}_d(\bm{\theta}_d^t, \bm{\tilde{\theta}}_d^t) \rVert^2 \leq \frac{\mathbb{E}[\mathcal{F}_d(\bm{\theta}_d^0, \bm{\tilde{\theta}}_d^0)-\mathcal{F}_d(\bm{\theta}_d^T, \bm{\tilde{\theta}}_d^T)]}{\alpha\gamma_d\epsilon TH} \\
        + \mathcal{O}\left(\frac{1}{\epsilon}\left(\gamma_d H^2 + \alpha D + \alpha^2\gamma_d D^2 H + \gamma_d D^2H\right)\right)
    \end{split}
\end{equation}
\end{theorem}

\subsection{Server-side model convergence}

The server-side model is trained in a centralized way, while using activations that generated by the device-side model instead of the raw data. The density of the activation distribution $\mathcal{A}$ is denoted as $p^t(\xi)$, which changes with the device-side model. Let $q^t = \int | p^t(\xi) - p^*(\xi) |d\xi$, where $p^*(\xi)$ is the converged density of activations that computed from converged device-side model. It has been proved in literature~\cite{decoupled} that layers trained on activations converges if all the previous layers converges.

Specifically, since the device-side model converges as presented in Equation~\ref{eq:device-converge}, according to Assumption 1, 3 and 4 and Proposition 3.1 presented in literature~\cite{decoupled}, we directly have

\begin{theorem}[Server Block Convergence]
\label{theorem:fedoptima-server-converge}
Each term of the following equation converges.
\begin{equation}
    \sum_{t=0}^T \gamma_s^t \mathbb{E} \lVert \nabla \mathcal{F}_s(\bm{\theta}_s^t) \rVert^2 \leq \mathbb{E}[\mathcal{F}_s(\bm{\theta}_s^0)] + G \sum_{t=0}^T \gamma_s^t \left(\sqrt{2q^t}+\frac{L\gamma_s^t}{2}\right)
\end{equation}
\end{theorem}

Thus the server-side model converges as presented in the literature~\cite{doi:10.1137/16M1080173,NEURIPS2018_a36b598a}.

\section{Implementation and setup}
\label{sec:implement}
The implementation of \MethodName\ is considered in Section~\ref{subsec:implement} and the experimental setup, including the testbed, models, datasets and baselines are presented in Section~\ref{subsec:setup}. 

\subsection{Implementation}
\label{subsec:implement}

\MethodName\ is developed using PyTorch to facilitate DNN partitioning. TCP/IP is used in server and device communication shown in Figure~\ref{subfig:training} to prevent packet loss.

\MethodName\ uses concurrent threads to implement the modules/submodules on the server and device. Four threads are created on the \textbf{\textit{server}}: for the Sender, Receiver, Task Scheduler, and Compute Engine. The Sender and Receiver connect to all devices. The Compute Engine places messages that are tagged with a destination into the sending queue of the Sender. The Sender checks the queue and sends messages to the intended device. The Receiver continuously collects messages from devices and puts them in a receiving queue. The Task Scheduler monitors the receiving queue and categorizes incoming messages and directs them to either the model or activation queues. The Compute Engine retrieves models and activations from the Task Scheduler used for aggregation and training.

Each \textit{\textbf{device}} has three threads: for the Sender, Receiver, and Compute Engine. The Sender and Receiver establish the connection with the server. The Compute Engine executes local training and puts activations into the sending queue. At the end of each training iteration, the Compute Engine places the local model in the sending queue and retrieves the global model from the receiving queue to complete aggregation.

\subsection{Experimental setup}
\label{subsec:setup}

\textbf{Testbeds}: 
\MethodName\ is evaluated on two distinct testbeds: 
(1) Testbed A consists of a CPU server and 8 Raspberry Pis with no hardware acceleration. This represents a resource-constrained FL system with edge devices, such as low-end smartphones. (2) Testbed B includes a GPU server and 16 Jetson Nanos with GPU-acceleration representative of high-end smartphones. The devices used are akin to those used in cross-device FL systems presented in the literature~\cite{10.1145/3485730.3493444,9762360,fedadapt}. 

The frequency of the device CPU/GPU is adjusted to achieve computational heterogeneity. The devices are divided into four groups, and devices within each group have the same compute and memory resources. The technical specification is provided in Table~\ref{tab:testbed}. The bandwidth between the server and the devices is 50 Mbps and 100 Mbps for Testbed~A and Testbed~B, respectively, unless stated otherwise.

\begin{table*}[tp]
    \centering
    \caption{Specification of the testbed. *Frequency adjusted to achieve computational heterogeneity.}
    \begin{tabular}{ccccccc}
        \Xhline{2\arrayrulewidth}
           \textbf{Testbed} & \textbf{} & \textbf{Group} & \textbf{Platform} & \textbf{CPU} & \textbf{GPU} & \textbf{Memory}  \\
        \Xhline{2\arrayrulewidth}
          \multirow{5}{*}{A} & Server &  & Dell Laptop & Intel i7-11850H & N/A & 16 GB \\
          \cline{2-7}
          & \multirow{4}{*}{Device} & a & Raspberry Pi 3B $\times$ 2 & Broadcom BCM2837 600 MHz* & N/A & 1 GB \\
          &  & b & Raspberry Pi 3B $\times$ 2 & Broadcom BCM2837 1.2 GHz & N/A & 1 GB \\
          &  & c & Raspberry Pi 4B $\times$ 2 & Broadcom BCM2711 1.2 GHz* & N/A & 4 GB \\
          &  & d & Raspberry Pi 4B $\times$ 2 & Broadcom BCM2711 1.8 GHz & N/A & 4 GB \\
          \Xhline{2\arrayrulewidth}
          \multirow{5}{*}{B} & Server &  & Workstation & AMD EPYC 7713p & NVIDIA RTX A6000 & 16 GB \\
          \cline{2-7}
           & \multirow{4}{*}{Device} & a & Jeston Nano $\times$ 4 & ARM Cortex-A57 1.7 GHz & NVIDIA GM20B 240 MHz* & 4 GB \\
          & & b & Jeston Nano $\times$ 4 & ARM Cortex-A57 1.7 GHz & NVIDIA GM20B 320 MHz* & 4 GB \\
          & & c & Jeston Nano $\times$ 4 & ARM Cortex-A57 1.7 GHz & NVIDIA GM20B 640 MHz* & 4 GB \\
          & & d & Jeston Nano $\times$ 4 & ARM Cortex-A57 1.7 GHz & NVIDIA GM20B 921 MHz & 4 GB \\
        \Xhline{2\arrayrulewidth}
    \end{tabular}
    \label{tab:testbed}
\end{table*}

\textbf{Tasks, DNNs and Datasets}: 
Two tasks, namely image classification and sentiment analysis, are considered.

For \textit{image classification}, Testbed~A uses a 5-layer VGG~\cite{vgg} (VGG-5) and CIFAR-10 dataset~\cite{cifar10,cifar10-2} that consists of 50,000 images with 10 classes for training and 10,000 images for testing. On Testbed~B, the MobileNetV3-Large~\cite{mobilenet} for mobile vision apps and the Tiny ImageNet dataset~\cite{tinyimagenet} are chosen. The training and test datasets contain 100,000 and 10,000 images, respectively, with 200 classes.

For sentiment analysis on Testbed A, a transformer~\cite{transformer} (Transformer-6) that consists of 6 transformer encoder layers is trained on the SST-2 dataset~\cite{sst2}, a standard dataset for sentiment analysis comprising 67,349 movie reviews (48,000 for training and 19,349 for testing) with binary sentiment labels. On Testbed B, a larger 12-layer transformer~\cite{transformer} (Transformer-12) is used to recognize complex linguistic patterns. The dataset is IMDB~\cite{imdb} that contains movie reviews with longer sentences. The training and test sets contain 25,000 samples.

The architectures of the DNNs are shown in Table~\ref{table:model-arch}. For VGG-5, `CONV-A-B' represents a convolutional layer of A$\times$A kernel size and of B output channels. `FC-A' represents a fully connected layer with the output size A. For MobileNet-Large, `BNECK-A-B' denotes a bottleneck residual block that consists of an expansion layer, a convolutional layer with the kernel size A$\times$A and a projection layer with the output channel number B. For both Transformer-6 and Transformer-12, `EMB-A' is a work embedding layer with output dimension A, and `ENC-A-B-C' is a transformer encoder layer with embedding dimension A, B heads and hidden dimension C. The number of classes is X. The activation, batch normalization and pooling layers are not shown for simplicity. 

The dataset is split in a non-IID manner across devices using the Dirichlet distribution with 0.5 prior based on the literature~\cite{acar2021federated}. Each device is assigned a vector with the size of the number of classes and its values are drawn from a Dirichlet distribution. These vectors represent the class distribution for each device. For each device, a label is randomly selected based on its corresponding vector, and a data point with this label is sampled without replacement. This continues until every data sample is allocated to a device.

\begin{table}[tp]
    \centering
    \caption{Neural network architectures chosen for this paper.}
    \begin{tabular}{ccc}
        \Xhline{2\arrayrulewidth}
        \textbf{VGG-5} & \textbf{MobileNetV3-Large} & \textbf{Transformer-6} \\
        \Xhline{2\arrayrulewidth}
        CONV-3-32 & CONV-3-16 & EMB-100 \\
        \hline
        \footnotesize{CONV-3-64 $\times$ 2} & BNECK-3-16 & \footnotesize{ENC-100-5-100 $\times$ 6} \\
        \hline
        FC-128 & BNECK-3-24 $\times$ 2 & FC-X \\
        \hline
        FC-X & BNECK-5-40 $\times$ 3 \\
        \cline{1-2}
         & BNECK-3-80 $\times$ 4 &  \\
        \cline{1-2}
         & BNECK-3-112 $\times$ 2 & \textbf{Transformer-12} \\
        \hline
         & BNECK-5-160 $\times$ 3 & EMB-100 \\
        \hline
         & CONV-1-960 & \footnotesize{ENC-100-50-100 $\times$ 12} \\
        \cline{2-3}
         & CONV-1-1280 & FC-X \\
        \cline{2-3}
         & FC-X & \\
        \Xhline{2\arrayrulewidth}
    \end{tabular}
    \label{table:model-arch}
\end{table}

\textbf{Baselines}: The performance of \MethodName\ is compared against six baselines: (a) classic FL~\cite{fedavg}, (b) two offloading-based FL methods, namely SplitFed~\cite{splitfed} and PiPar~\cite{pipar}, (c) two asynchronous FL methods, namely FedAsync~\cite{fedasync} and FedBuff~\cite{fedbuff}, and (d) a hybrid of offloading and asynchronous FL, OAFL (see Section~\ref{subsec:challenge} for details). 
The baselines are state-of-the-art methods and are further discussed in Section~\ref{sec:rw}.

\section{Results}
\label{sec:results}
The results demonstrate that \MethodName\ has superior training efficiency (Section~\ref{subsec:efficiency}), lower server and device idle time (Section~\ref{subsec:idle-time}), and higher throughput \di{in stable} (Section~\ref{subsec:throughput}) \di{and unstable (Section~\ref{subsec:throughput-unstable}) environments. We further present ablation studies (Section~\ref{subsec:ablation}) on auxiliary network design and task scheduling mechanisms to analyze the contribution of the key components of \MethodName.}

\subsection{Training efficiency}
\label{subsec:efficiency}

\begin{figure*}[tp]
    \centering
        \subfigure[Testbed A]{
	    \includegraphics[width=0.35\textwidth]{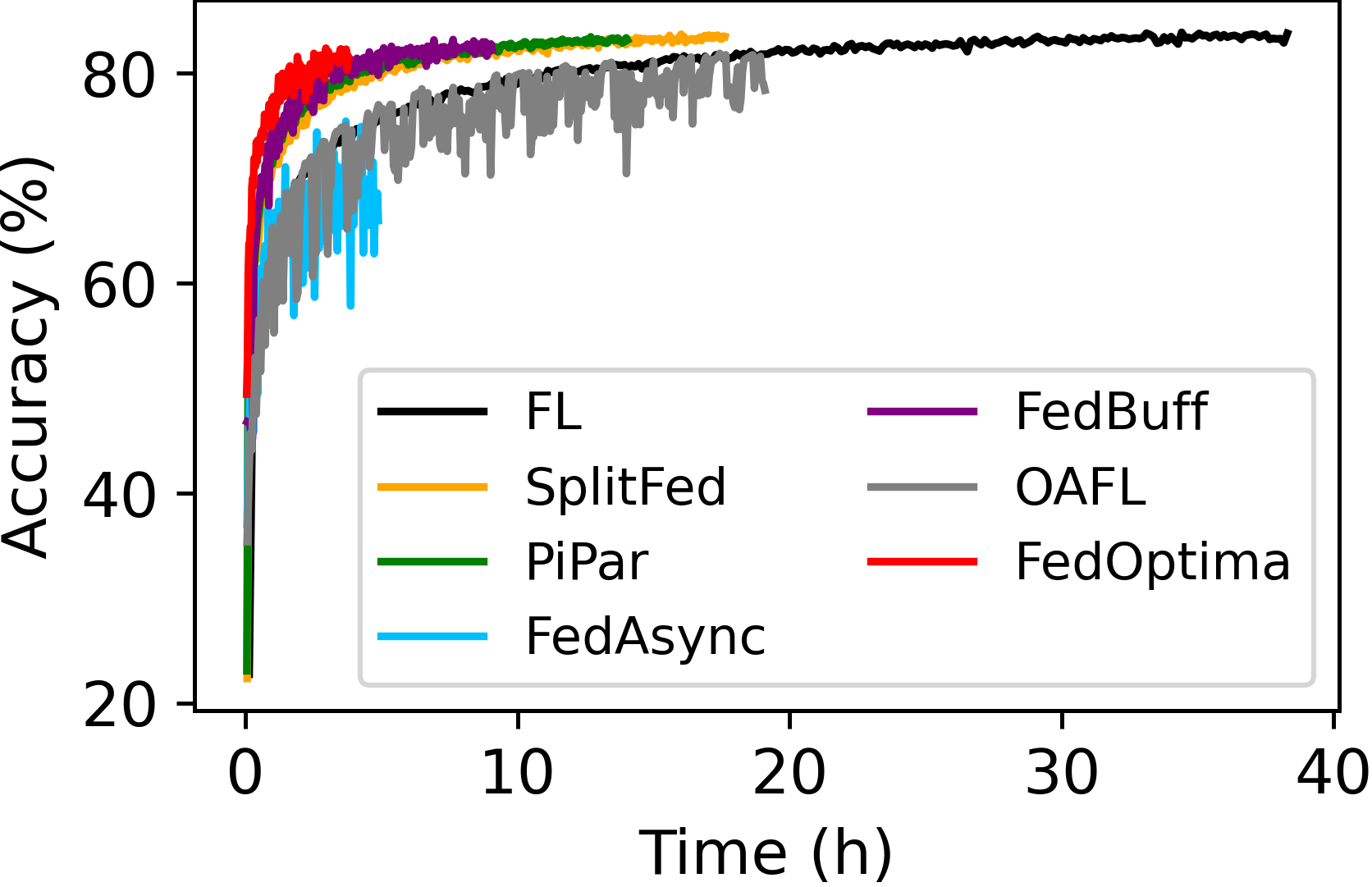}
	       \label{subfig:cpu_cv_acc}
	    }
        \subfigure[Testbed B]{
	    \includegraphics[width=0.35\textwidth]{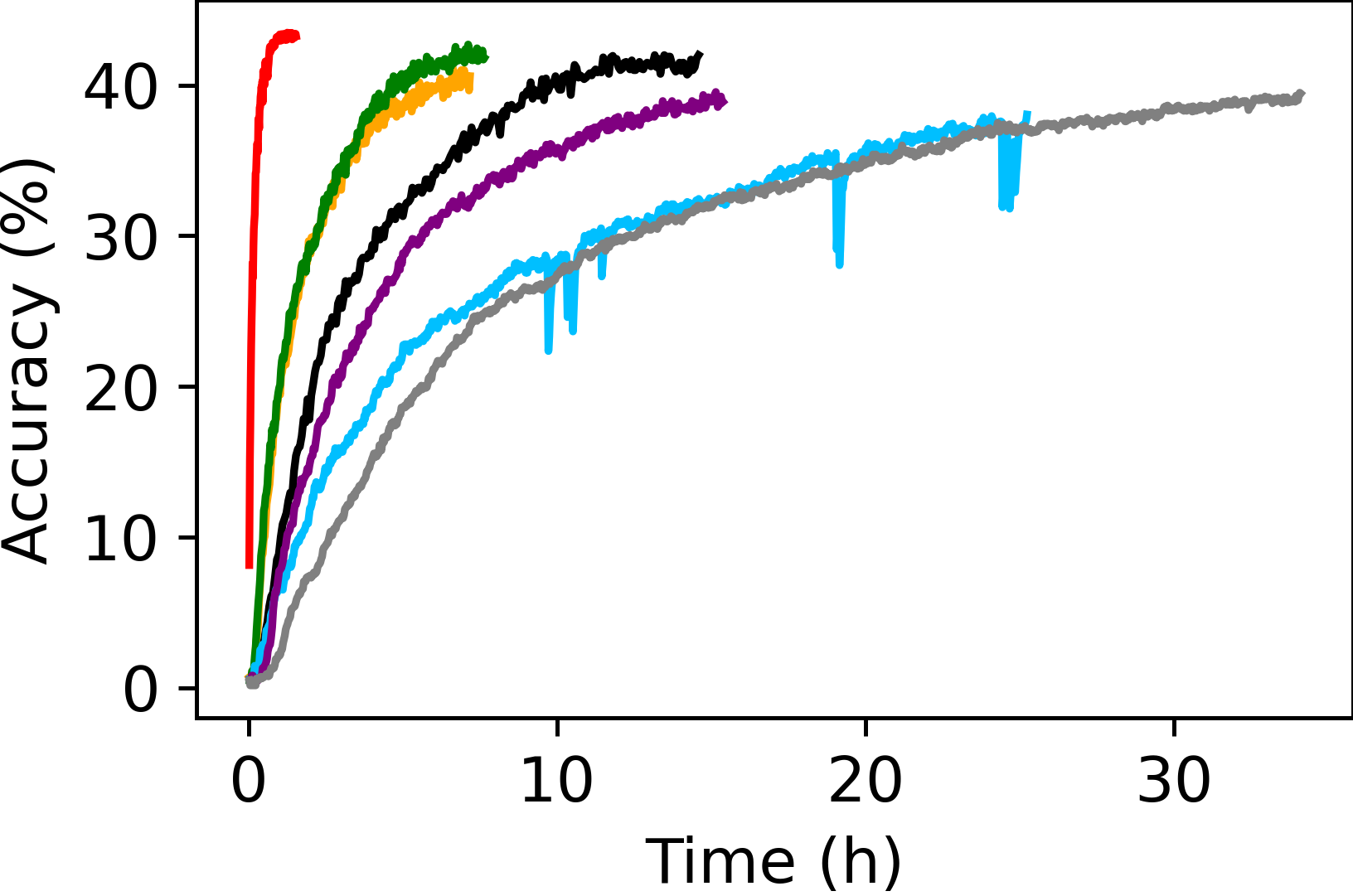}
	       \label{subfig:gpu_cv_acc}
	    }
	\caption{Accuracy (higher is better) versus training time (lower is better) for image classification.}
        \label{fig:train-time-cv}
\end{figure*}

\begin{figure*}[tp]
\centering
        \subfigure[Testbed A]{
	    \includegraphics[width=0.35\textwidth]{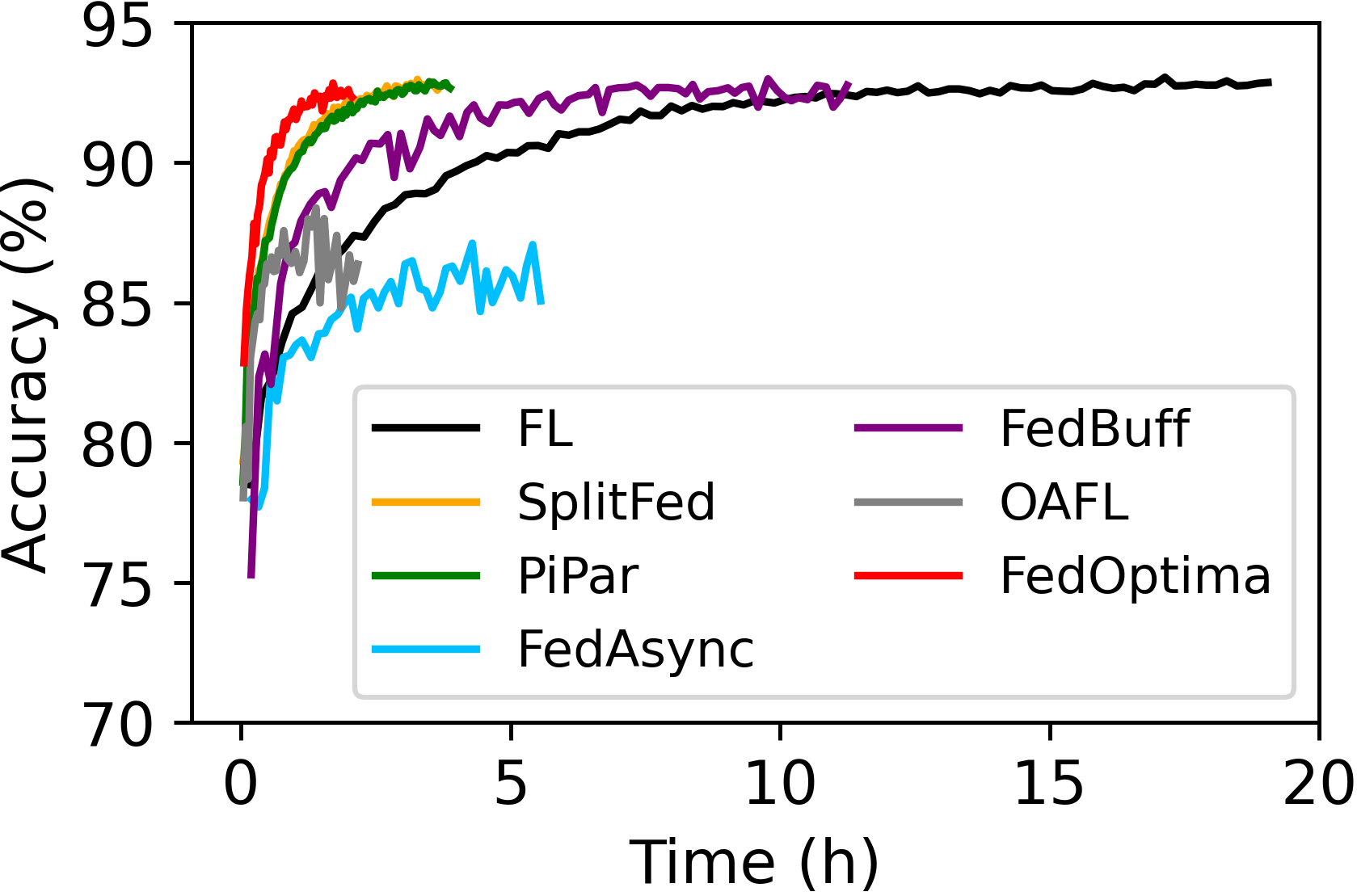}
	       \label{subfig:cpu_nlp_acc}
	    }
        \subfigure[Testbed B]{
	    \includegraphics[width=0.35\textwidth]{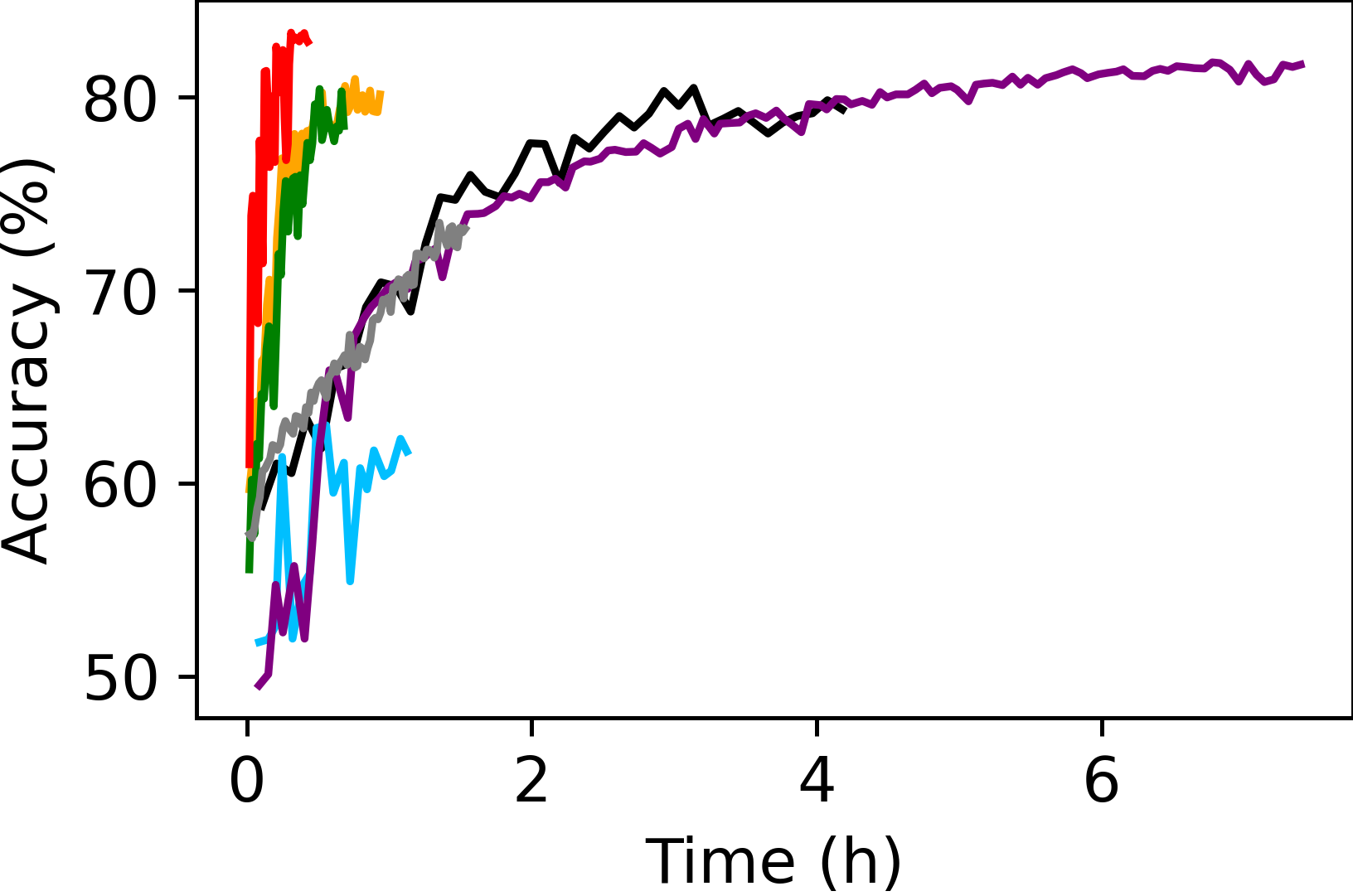}
	       \label{subfig:gpu_nlp_acc}
	    }
	\caption{Accuracy (higher is better) versus training time (lower is better) for sentiment analysis.}
        \label{fig:train-time-nlp}
\end{figure*}

End-to-end training for the two tasks, namely image classification and sentiment analysis, is considered for \MethodName\ and the baselines. The wall-clock training time, model convergence rate and final accuracy are measured. The accuracy curves are shown in Figure~\ref{fig:train-time-cv} and Figure~\ref{fig:train-time-nlp}. Each curve shows the validation accuracy versus time for model convergence. \di{We use validation-based early stopping~\cite{prechelt1998earlystopping} and consider a model to have converged when the validation accuracy does not improve for 30 consecutive epochs; this criterion is applied consistently to all methods.}

\di{Figure~\ref{fig:train-time-cv} shows that \MethodName\ accelerates convergence for image classification on both testbeds. On Testbed~A, \MethodName\ reaches 82.4\% accuracy in 2.95 hours. FedAsync and OAFL do not reach comparable accuracy, whereas SplitFed, PiPar, FedBuff, and FL achieve similar accuracy but require 5.4$\times$, 4.3$\times$, 2.6$\times$, and 11.7$\times$ more time, respectively. On Testbed~B, where more layers are trained on the server, the advantage of \MethodName\ is more pronounced: it maintains higher accuracy than all baselines throughout training. Only FL and PiPar achieve a comparable final accuracy (42.3\%), but they require 18.7$\times$ and 10.0$\times$ more time. Overall, \MethodName\ achieves the highest final accuracy among all baselines.}

\di{The same trend holds for sentiment analysis (Figure~\ref{fig:train-time-nlp}). On both testbeds, \MethodName\ converges faster than all baselines. On Testbed~A, \MethodName\ reaches 92.84\% accuracy in 1.71 hours; FedAsync, FedBuff, and OAFL do not reach this accuracy. SplitFed, PiPar, and FL achieve comparable final accuracy but require 1.9$\times$, 2.1$\times$, and 10.1$\times$ more time than \MethodName, respectively. On Testbed~B, \MethodName\ achieves the shortest convergence time (0.31 hours) and the highest final accuracy (83.4\%). FedBuff yields the next highest accuracy (81.8\%) but requires 21.8$\times$ more time than \MethodName.}

\begin{boxH}
    \textbf{Observation 1:} Models trained using \MethodName\ converge faster than all baselines, and achieve final accuracy that is higher or comparable to the best baseline.
\end{boxH}

\subsection{Idle time analysis}
\label{subsec:idle-time}

\begin{figure*}[tp]
    \centering
        \subfigure[Testbed A - Server Side]{
	    \includegraphics[width=0.35\linewidth]{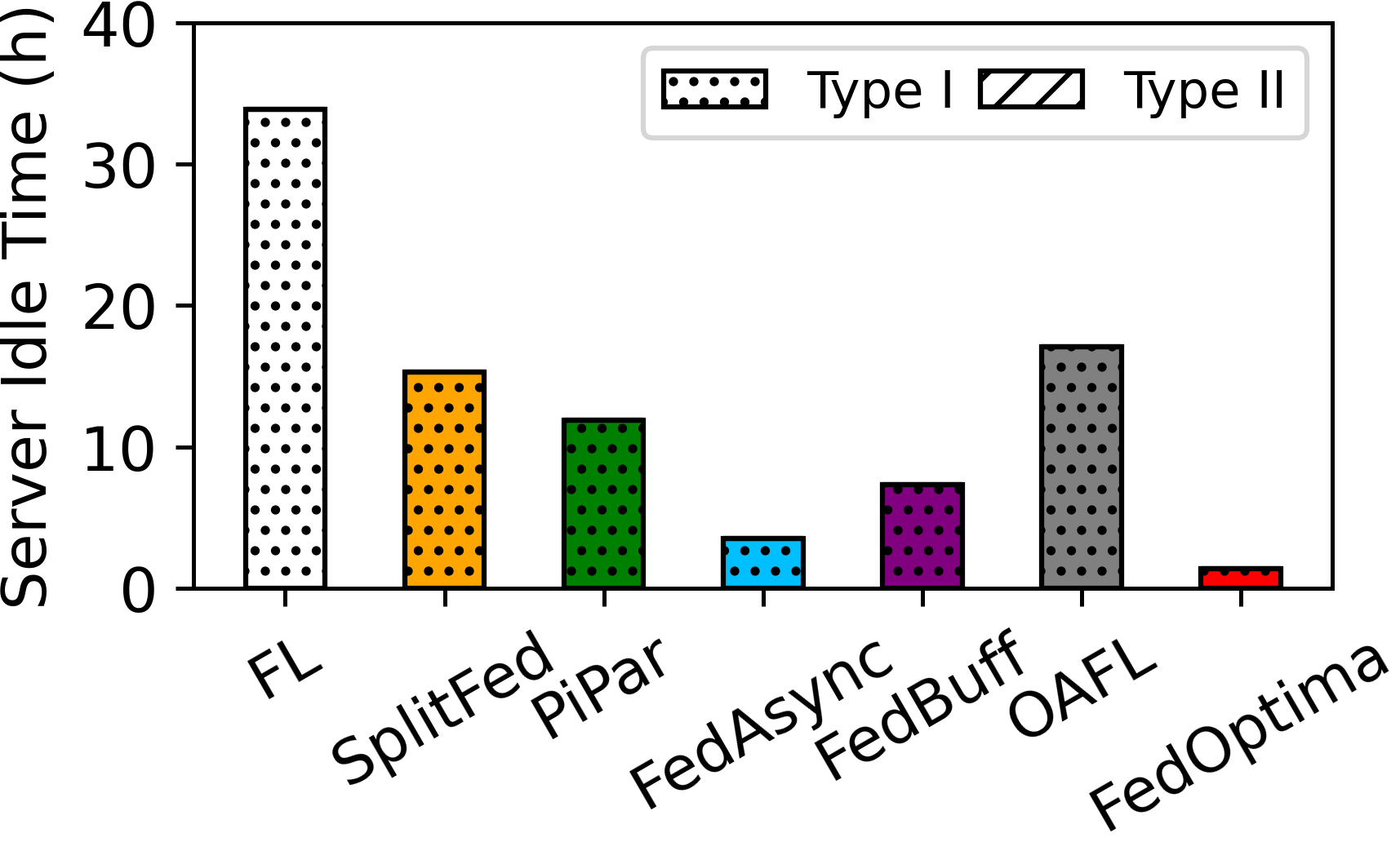}
	       \label{subfig:cpu_cv_server_idle}
	    }
        \subfigure[Testbed A - Device Side]{
	    \includegraphics[width=0.35\linewidth]{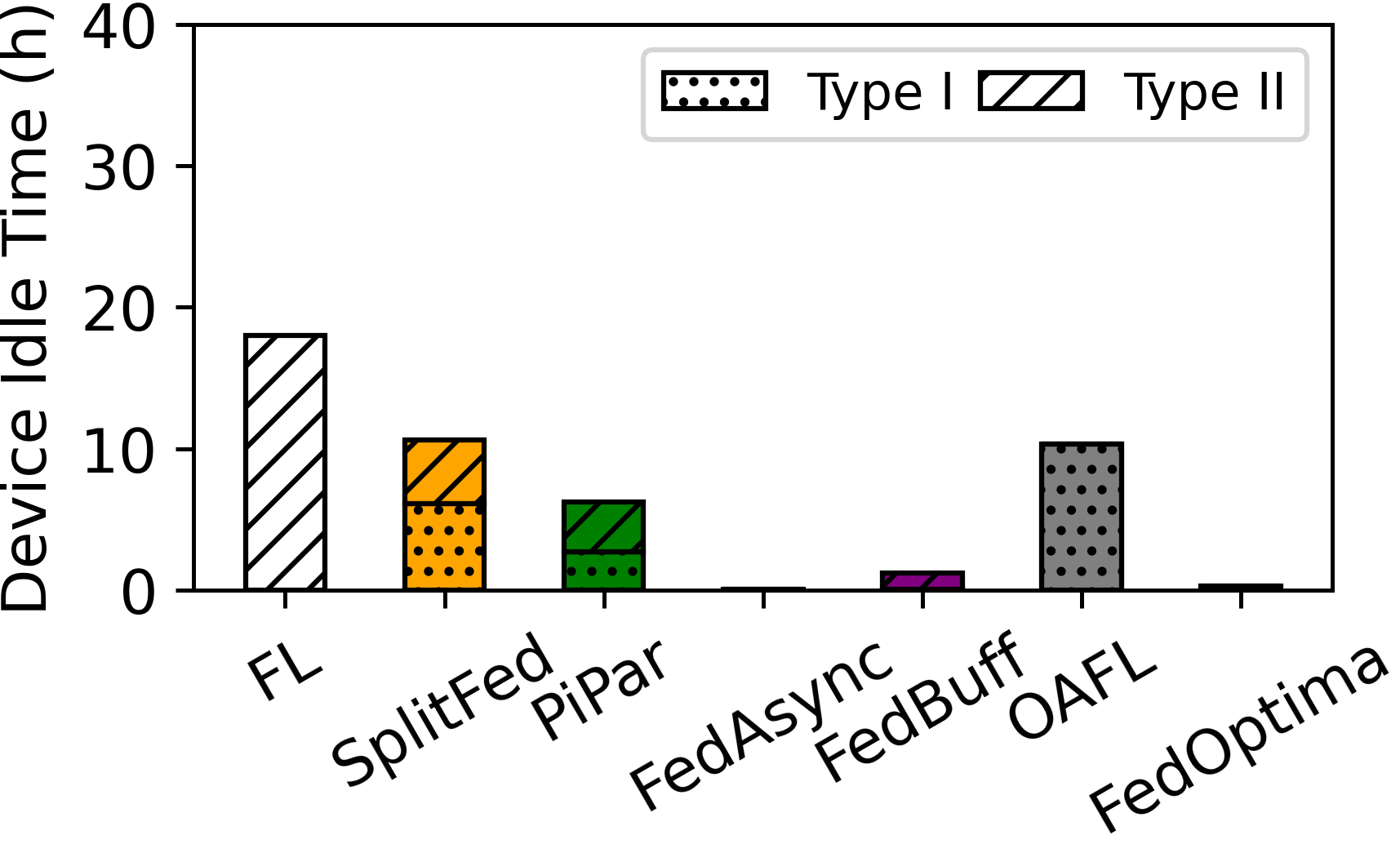}
	       \label{subfig:cpu_cv_device_idle}
	    }
        \subfigure[Testbed B - Server Side]{
	    \includegraphics[width=0.35\linewidth]{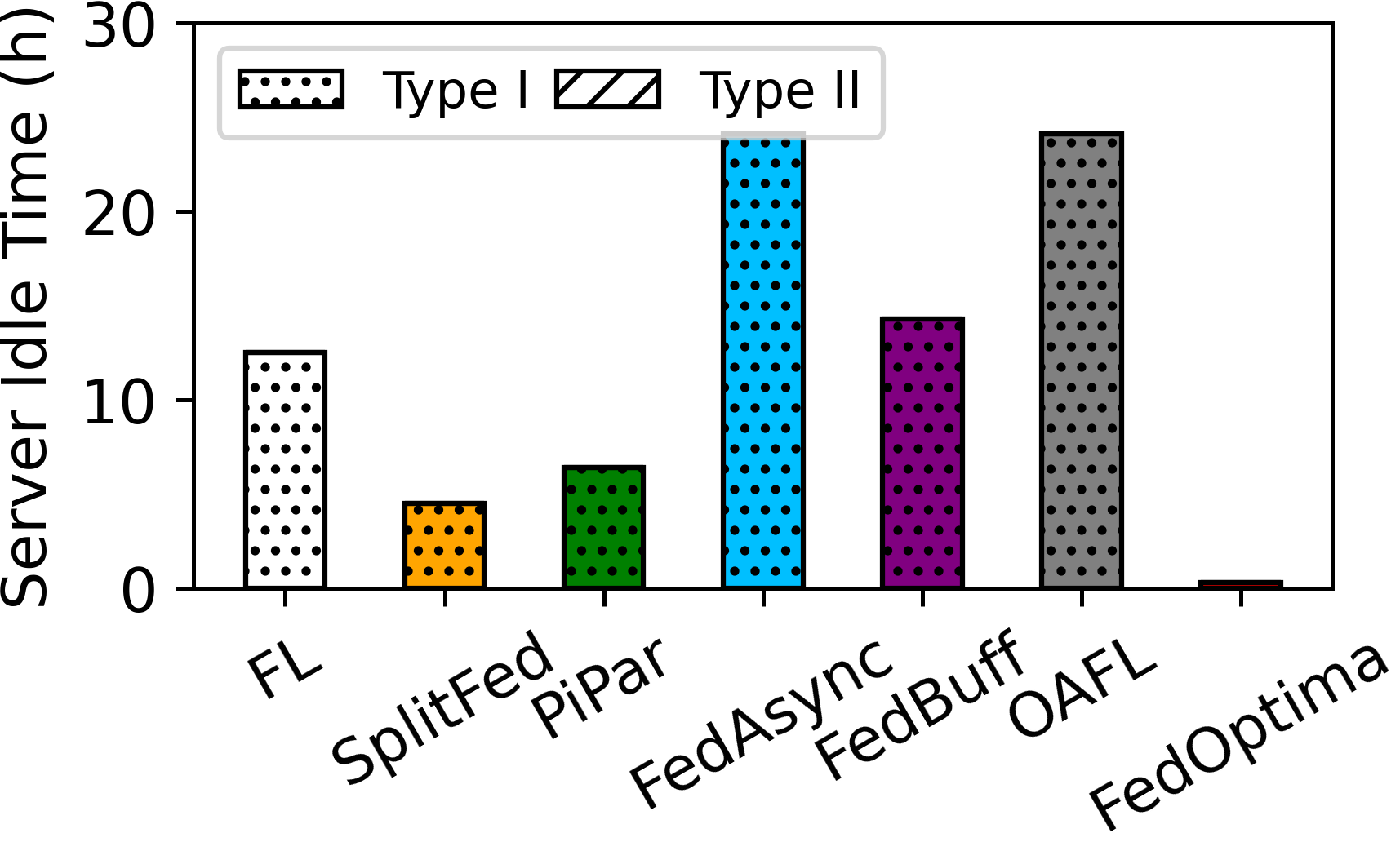}
	       \label{subfig:gpu_cv_server_idle}
	    }
        \subfigure[Testbed B - Device Side]{
	    \includegraphics[width=0.35\linewidth]{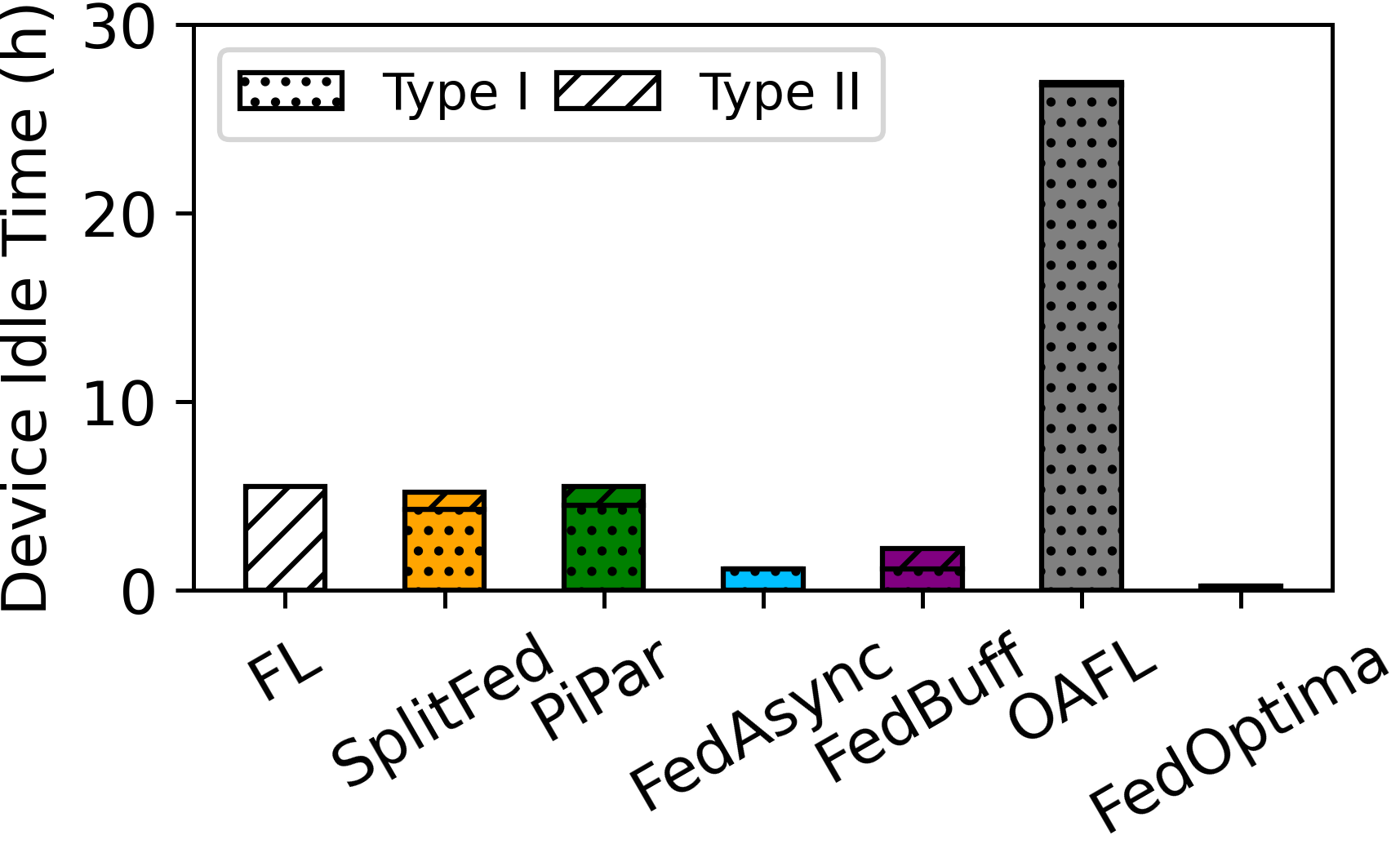}
	       \label{subfig:gpu_cv_device_idle}
	    }
	\caption{Server and device idle time (lower is better) of image classification.}
        \label{fig:idle-time-cv}
\end{figure*}

\begin{figure*}[tp]
    \centering
        \subfigure[Testbed A - Server Side]{
	    \includegraphics[width=0.35\linewidth]{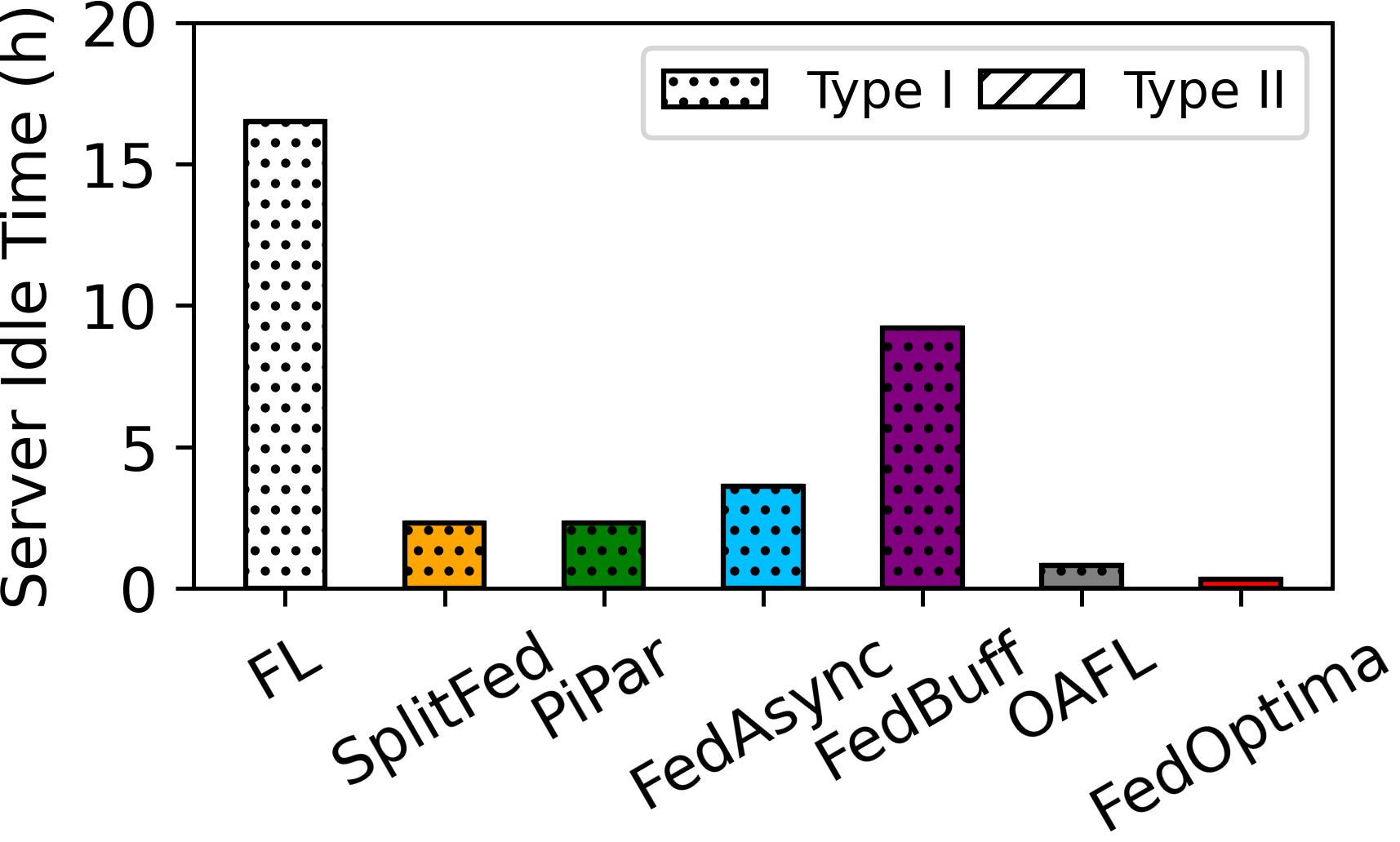}
	       \label{subfig:cpu_nlp_server_idle}
	    }
        \subfigure[Testbed A - Device Side]{
	    \includegraphics[width=0.35\linewidth]{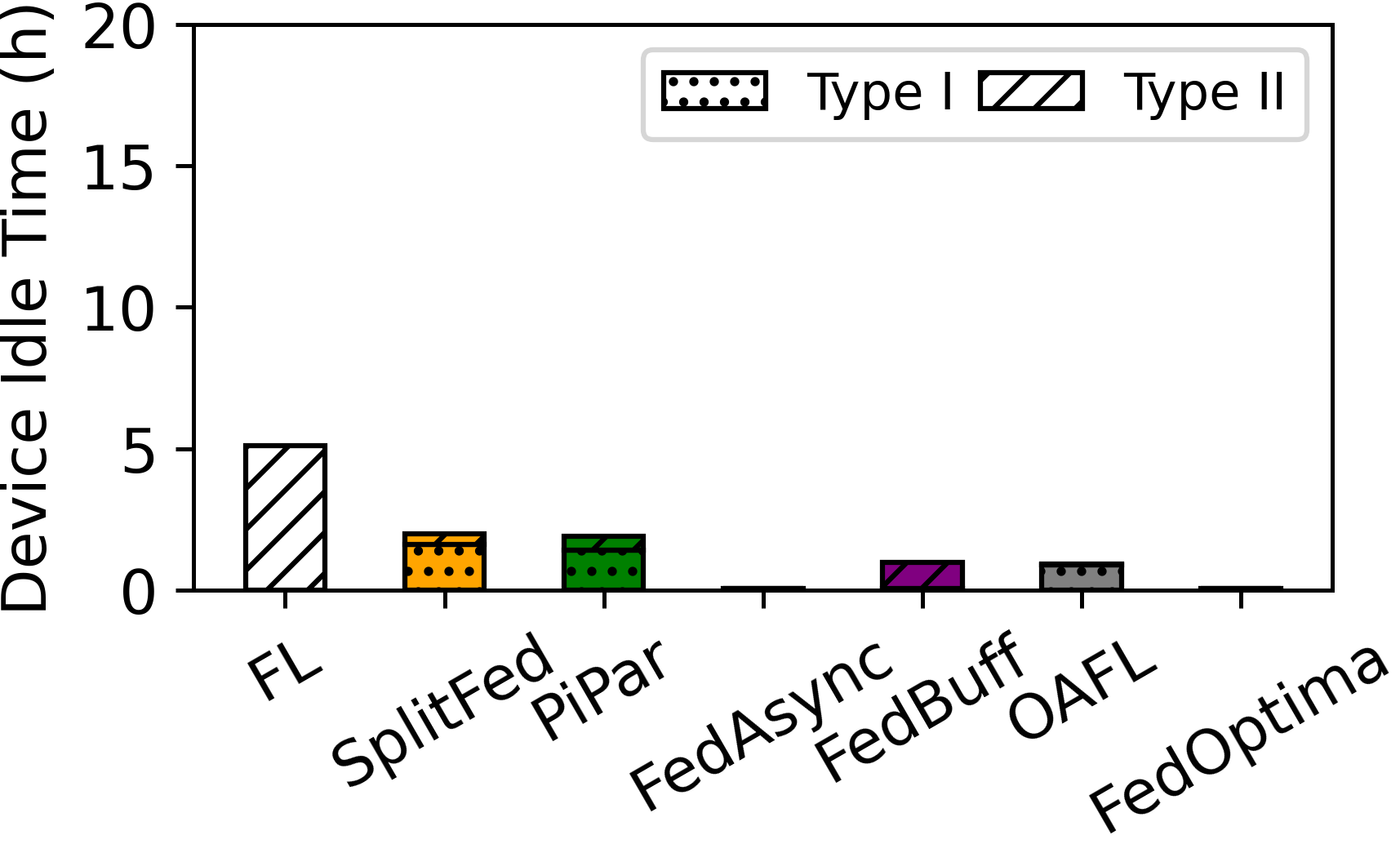}
	       \label{subfig:cpu_nlp_device_idle}
	    }
        \subfigure[Testbed B - Server Side]{
	    \includegraphics[width=0.35\linewidth]{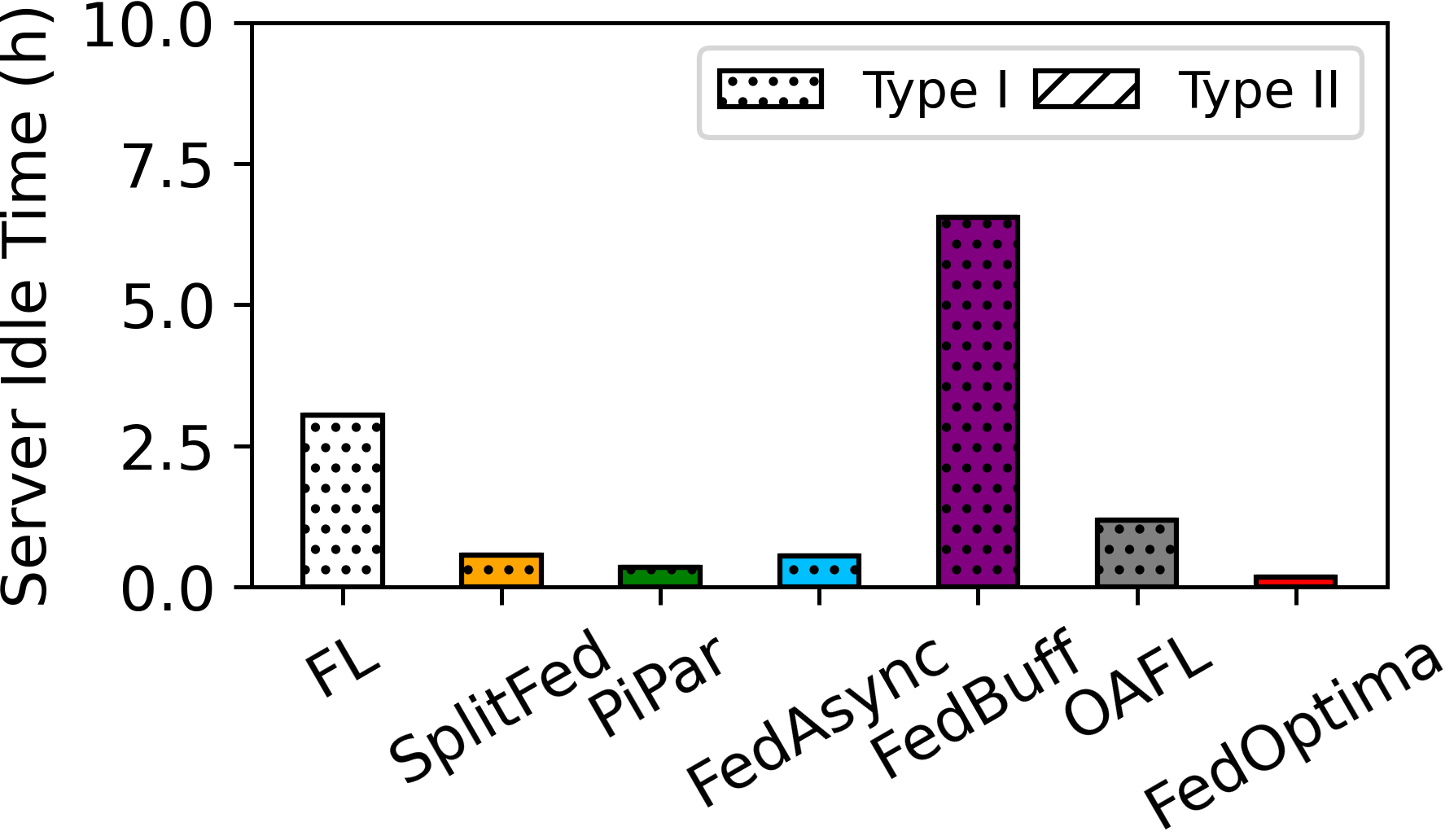}
	       \label{subfig:gpu_nlp_server_idle}
	    }
        \subfigure[Testbed B - Device Side]{
	    \includegraphics[width=0.35\linewidth]{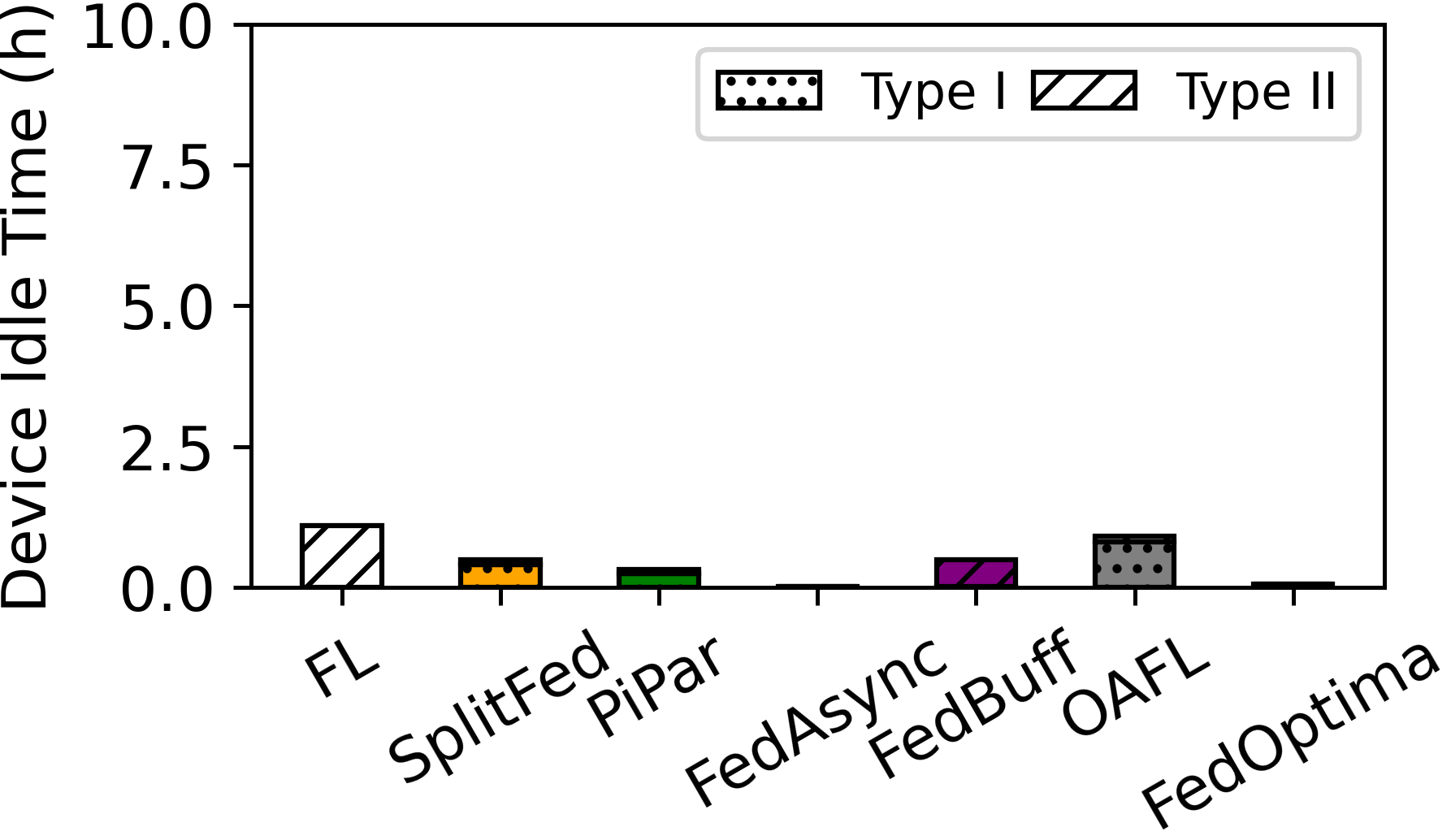}
	       \label{subfig:gpu_nlp_device_idle}
	    }
	\caption{Server and device idle time (lower is better) of sentiment analysis.}
        \label{fig:idle-time-nlp}
\end{figure*}

\di{Figure~\ref{fig:idle-time-cv} and Figure~\ref{fig:idle-time-nlp} shows the idle time for the two tasks. Device idle time is averaged across all devices. \MethodName\ achieves the lowest idle time in all cases. Only FedAsync has comparable device idle time to \MethodName, but it incurs a higher server idle time. Compared to the baseline with the lowest idle time, \MethodName\ reduces idle time by up to 93.9\% on the server and 81.8\% on devices.}

\begin{boxH}
    \textbf{Observation 2:} \MethodName\ has the lowest server and device idle time compared to all baselines.
\end{boxH}

\subsection{\di{System throughput in stable environments}}
\label{subsec:throughput}

\begin{figure*}[tp]
\centering
        \subfigure[Testbed A]{
	    \includegraphics[width=0.35\textwidth]{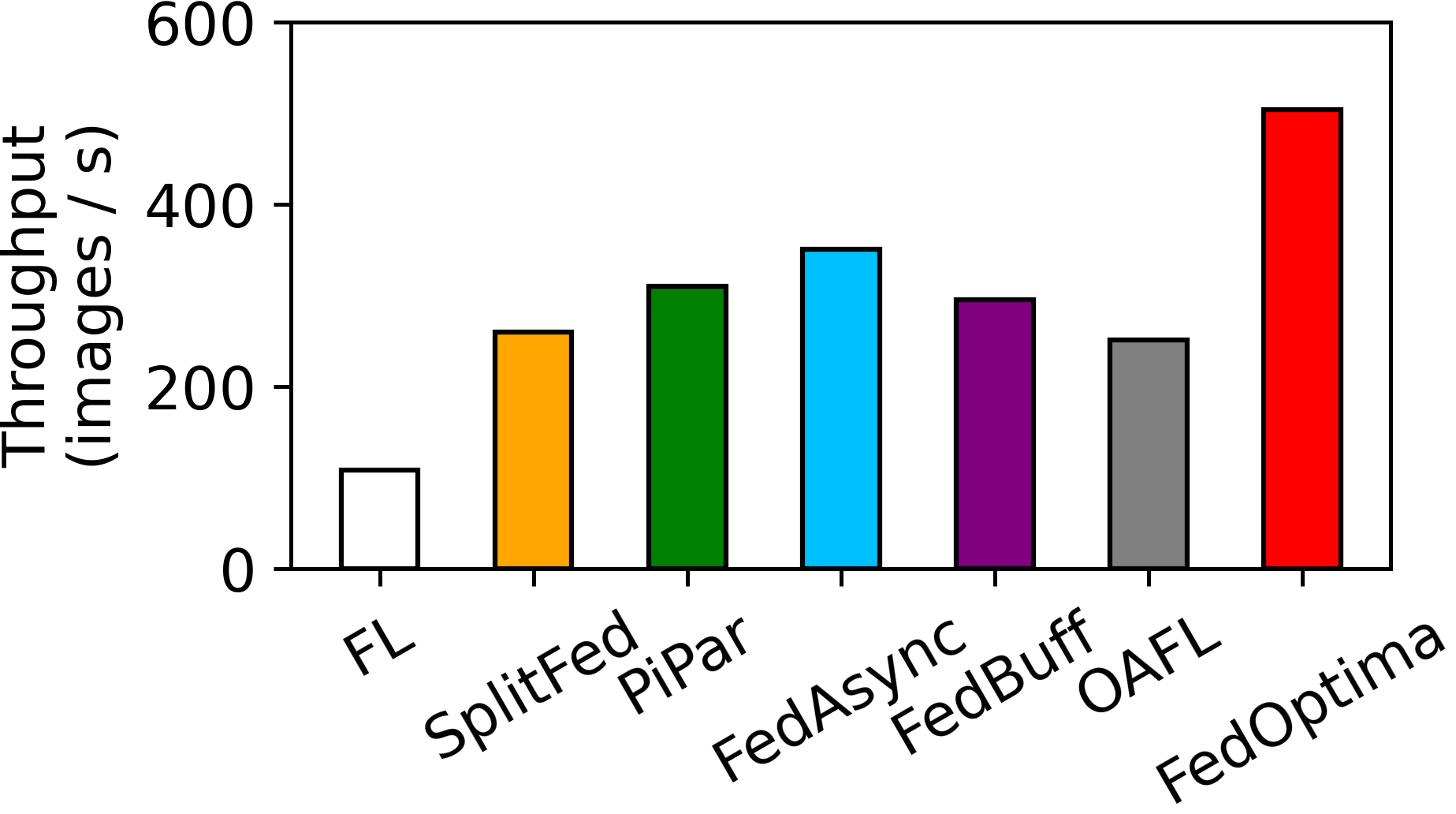}
	       \label{subfig:cpu_cv_throughput}
	    }
        \subfigure[Testbed B]{
	    \includegraphics[width=0.35\textwidth]{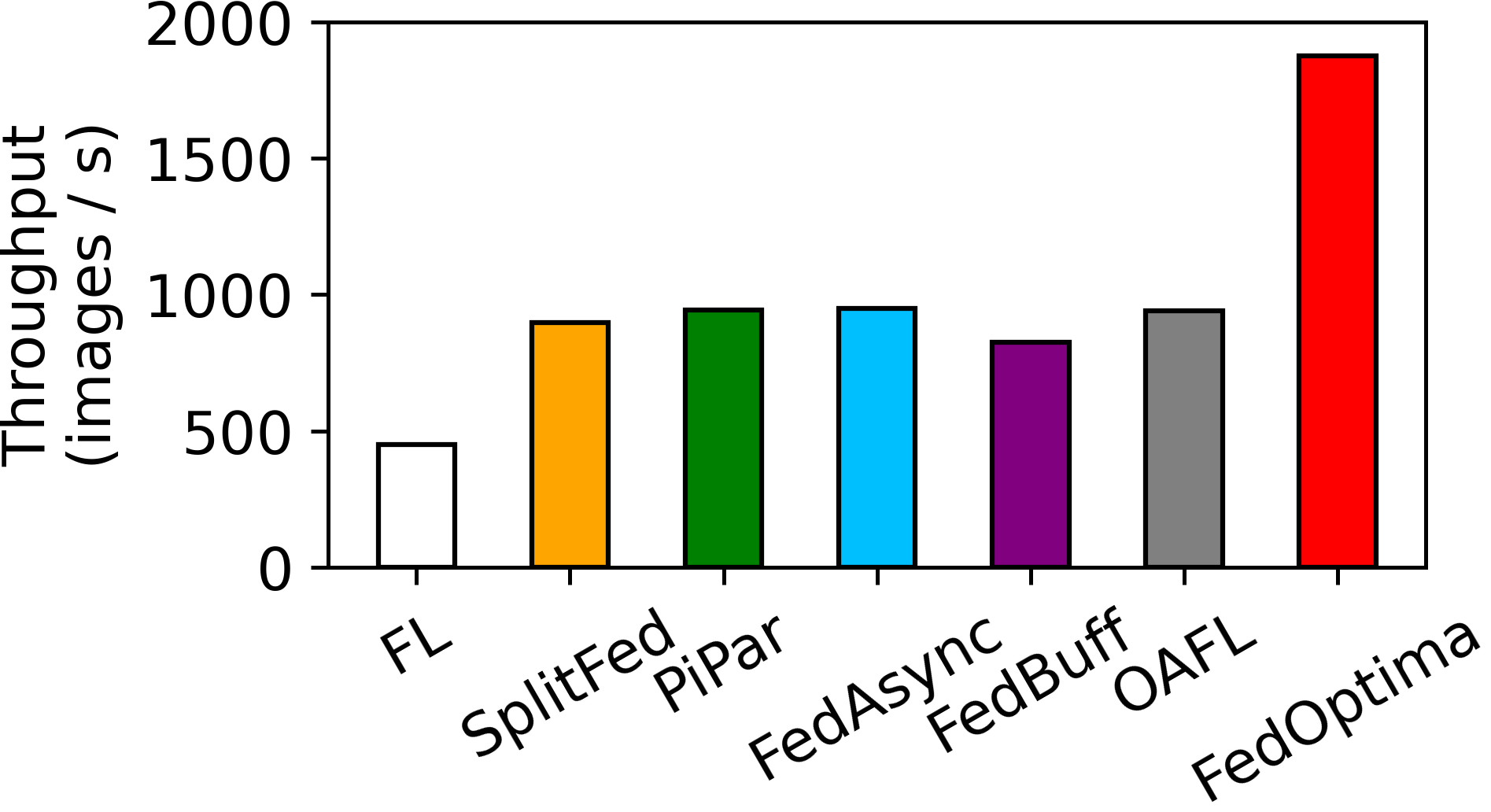}
	       \label{subfig:gpu_cv_throughput}
	    }
	\caption{System throughput (higher is better) for image classification.}
        \label{fig:throughput-cv}
\end{figure*}

\begin{figure*}[tp]
\centering
        \subfigure[Testbed A]{
	    \includegraphics[width=0.35\textwidth]{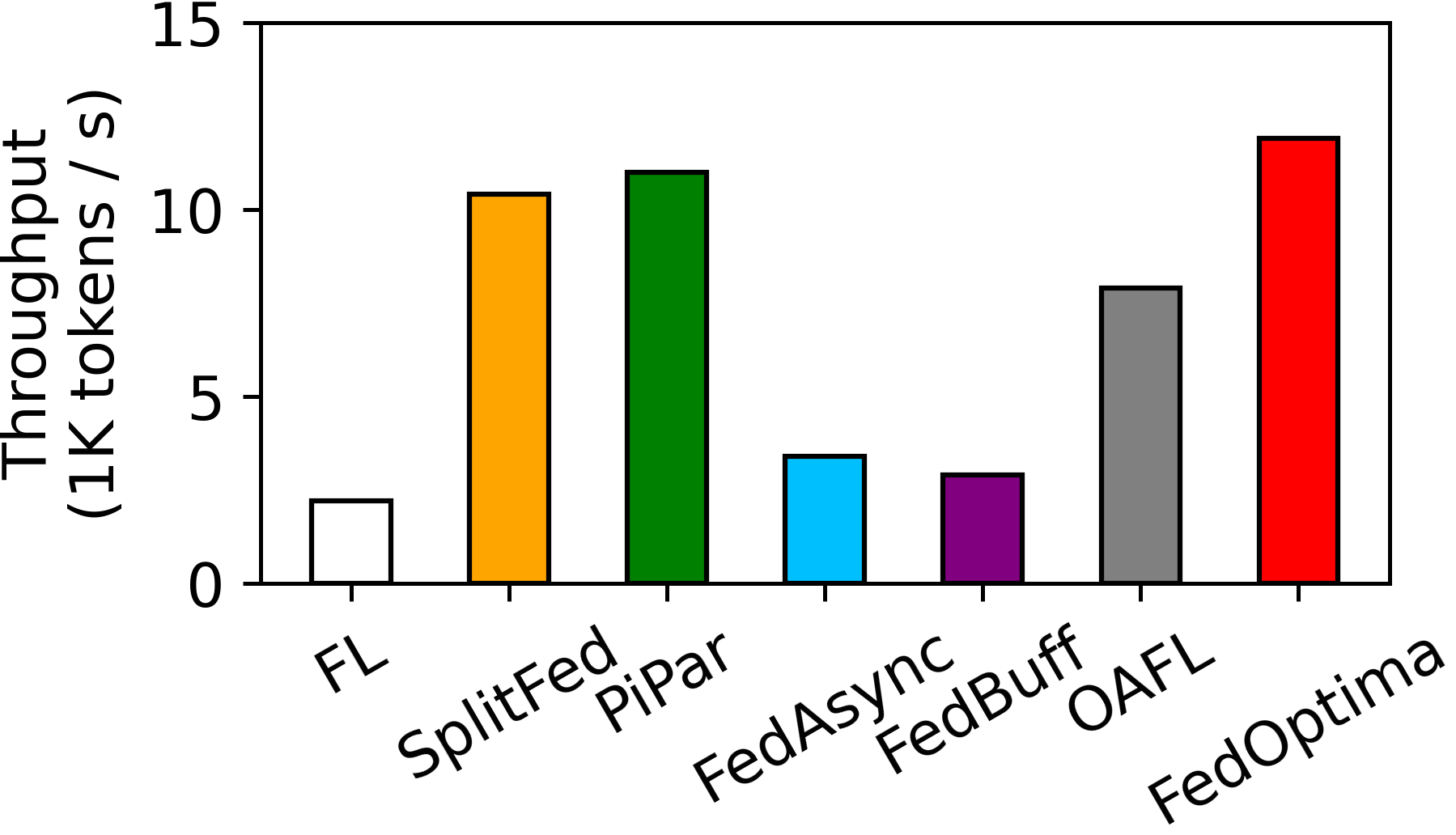}
	       \label{subfig:cpu_nlp_throughput}
	    }
        \subfigure[Testbed B]{
	    \includegraphics[width=0.35\textwidth]{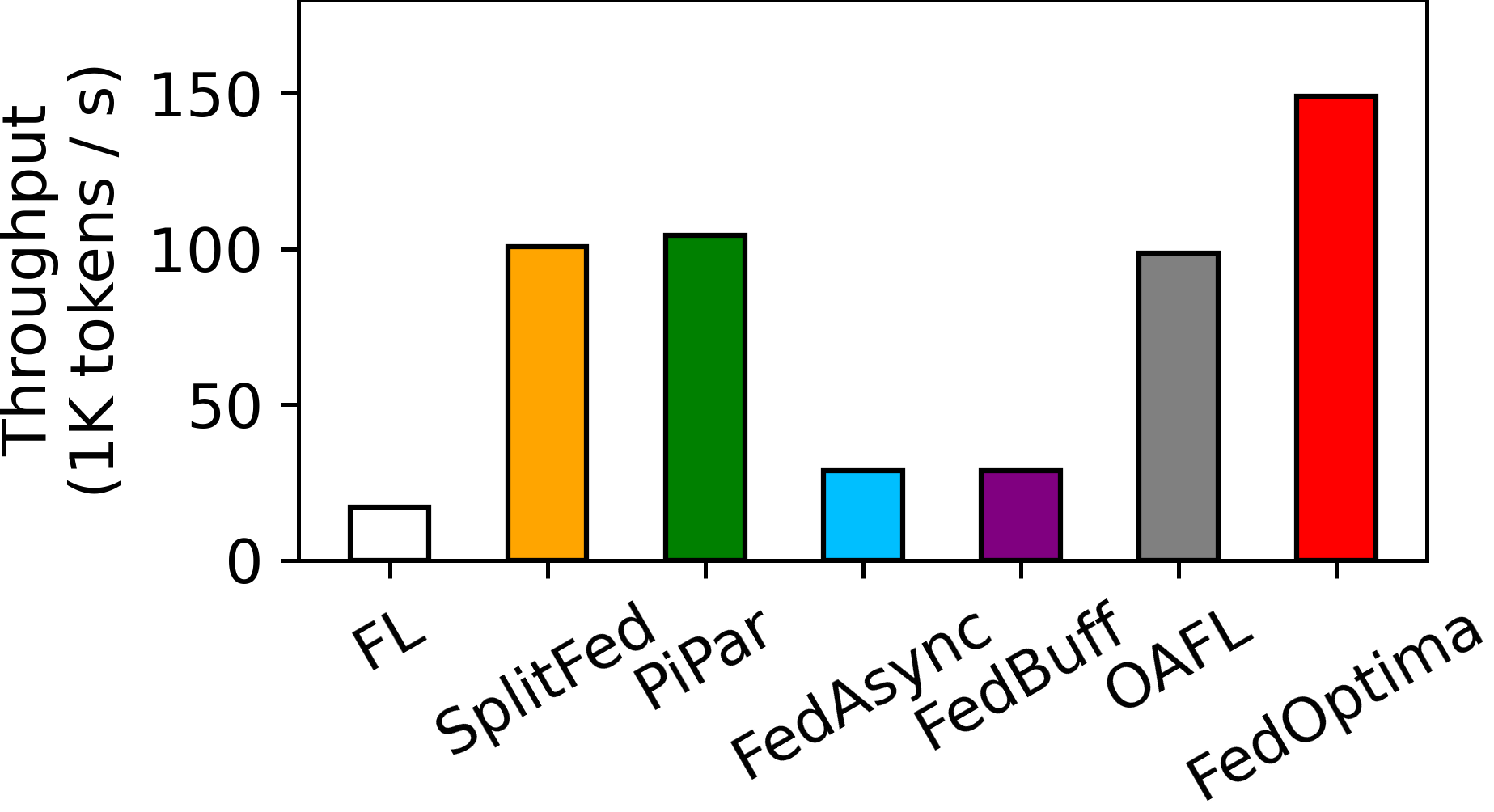}
	       \label{subfig:gpu_nlp_throughput}
	    }
	\caption{System throughput (higher is better) for sentiment analysis.}
        \label{fig:throughput-nlp}
\end{figure*}

\di{Throughput, which is the number of data samples processed per unit time is another indicator of system performance. In this section, we report throughput in stable environments (i.e. all devices participate throughout, and the available bandwidth between them and the server remains constant.). Throughput resilience under unstable environments in which the bandwidth varies and devices dropout is evaluated in Section~\ref{subsec:throughput-unstable}.}

\di{In stable environments, throughput reflects how effectively a method overlaps computation and communication and avoids stragglers or blocking behaviors. Therefore, higher throughput is due to better resource utilization (i.e., lower server/device idle time as observed in Section~\ref{subsec:idle-time}).}

Figure~\ref{fig:throughput-cv} reports image classification throughput. \MethodName\ achieves the highest throughput on both testbeds. On Testbed~A, FedAsync reaches 70\% of \MethodName's throughput, while other baselines are lower. On Testbed~B, \MethodName\ achieves 2.0$\times$ higher throughput than the next best method (FedAsync). For sentiment analysis (Figure~\ref{fig:throughput-nlp}), \MethodName\ again achieves the highest throughput: PiPar reaches 92\% of \MethodName's throughput on Testbed~A, whereas on Testbed~B \MethodName\ outperforms PiPar by 1.4$\times$.

\di{The above results show the throughput for \MethodName\ and the baselines in a stable environment, which is used in the next section to quantify throughput under bandwidth variation and device dropout.}

\begin{boxH}
    \textbf{Observation 3:} \di{\MethodName\ achieves the highest throughput compared to all baselines.}
\end{boxH}

\subsection{\di{Throughput resilience in unstable environments}}
\label{subsec:throughput-unstable}

\di{In practical FL deployments, devices may drop out (device churn) and network bandwidth can fluctuate over time, which can reduce effective system throughput. To evaluate the throughput resilience of \MethodName\ under such unstable conditions, we emulate bandwidth variation and device dropout, and measure how throughput degrades as the dropout probability increases.}

\di{\textbf{Modeling an unstable environment}:}
\di{Throughput is affected by unstable networks in which the bandwidth between the server and devices varies and devices may join or leave. We emulate this setting by periodically and randomly adjusting bandwidth and adding/removing devices. The network condition for each device changes every 10 minutes. At each bandwidth update, each device independently drops out of training with probability $p$ (with $0 \leq p \leq 0.5$). Devices that drop out do not participate in training during the subsequent interval, and they may rejoin at the next bandwidth update, again following the same dropout probability $p$. For devices that remain connected, the bandwidth is randomly reset to emulate the change of bandwidth. Bandwidth is uniformly sampled from 25--50~Mbps for Testbed~A and 50--100~Mbps for Testbed~B, which represent moderate bandwidth conditions (e.g., typical mobile or home Wi‑Fi links) and preserve a clear separation between the two testbeds.}

\di{\textbf{Baselines}:}
\di{We use FedAsync (image classification) and PiPar (sentiment analysis) as representative baselines because they achieve the highest throughput among the baselines in the stable setting; this makes the comparison under churn and bandwidth variation more stringent.}

\di{\textbf{Resilience metric}:
To quantify throughput resilience under device churn, we define the \emph{retention ratio} as $R(p)=T(p)/T(0)$, where $T(p)$ denotes throughput under dropout probability $p$, and $T(0)$ is the throughput when no device drops out. Here, \emph{retention} is the fraction of throughput in the stable environment that is preserved under instability; this ratio normalizes the throughput under instability of each method by its stable-environment baseline. For example, $R(p)=1$ means no degradation and smaller values indicate larger relative throughput loss as the environment becomes more unstable.}

\begin{figure*}[tp]
\centering
        \subfigure[Testbed A]{
	    \includegraphics[width=0.35\textwidth]{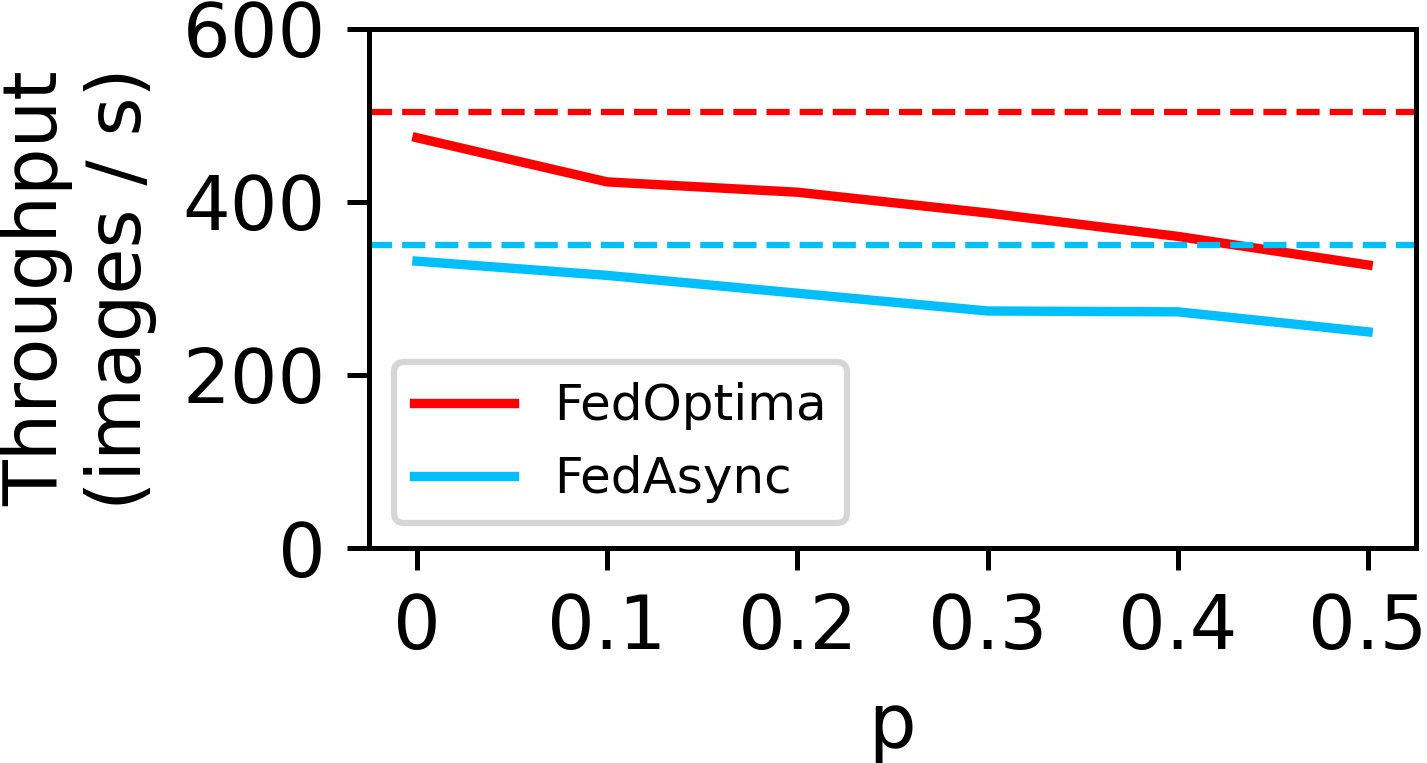}
	       \label{subfig:cpu_cv_throughput_unstable}
	    }
        \subfigure[Testbed B]{
	    \includegraphics[width=0.35\textwidth]{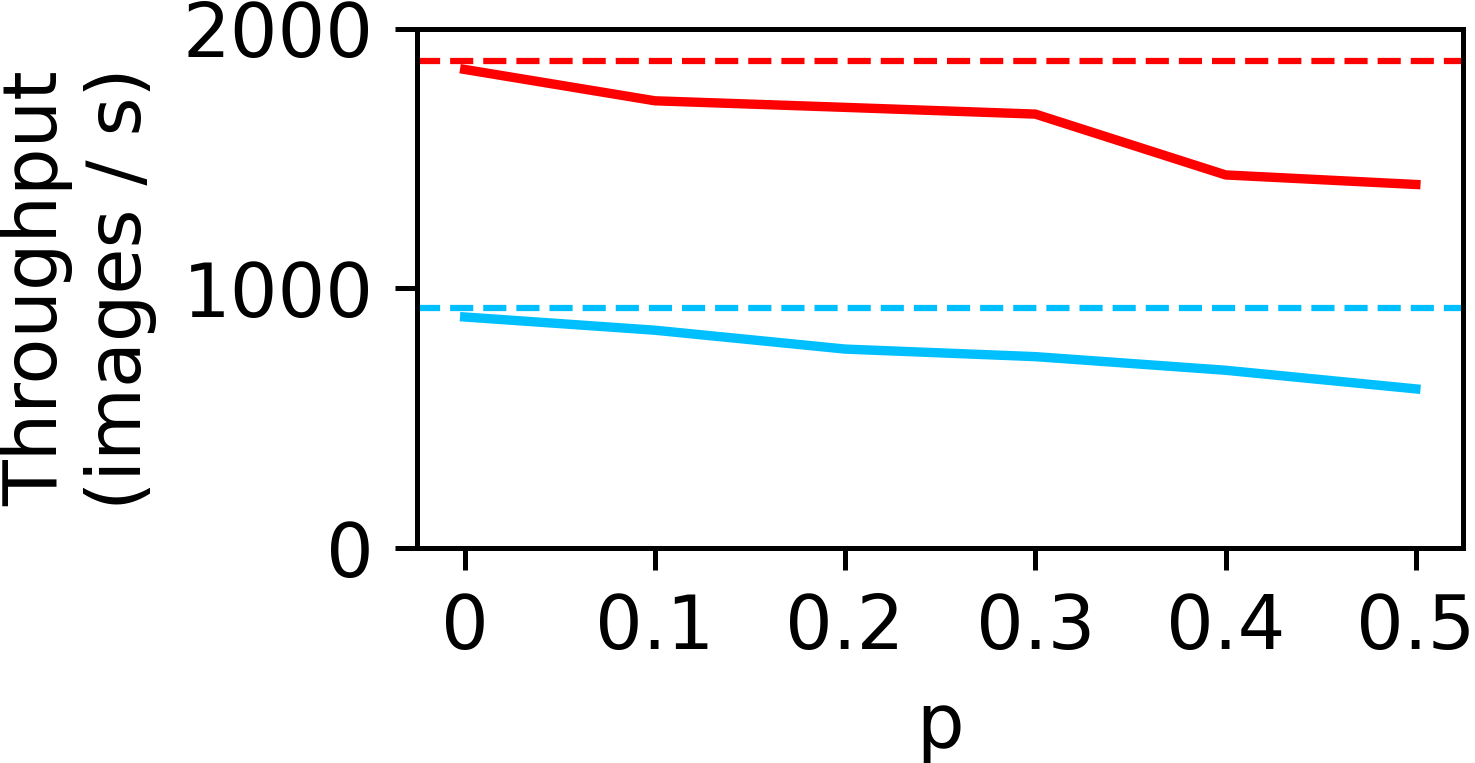}
	       \label{subfig:gpu_cv_throughput_unstable}
	    }
	\caption{\di{Throughput (higher is better) in an unstable network environment with different probabilities of device leaving ($p$) for image classification.}}
        \label{fig:throughput-cv-unstable}
\end{figure*}

\begin{figure*}[tp]
\centering
        \subfigure[Testbed A]{
	    \includegraphics[width=0.35\textwidth]{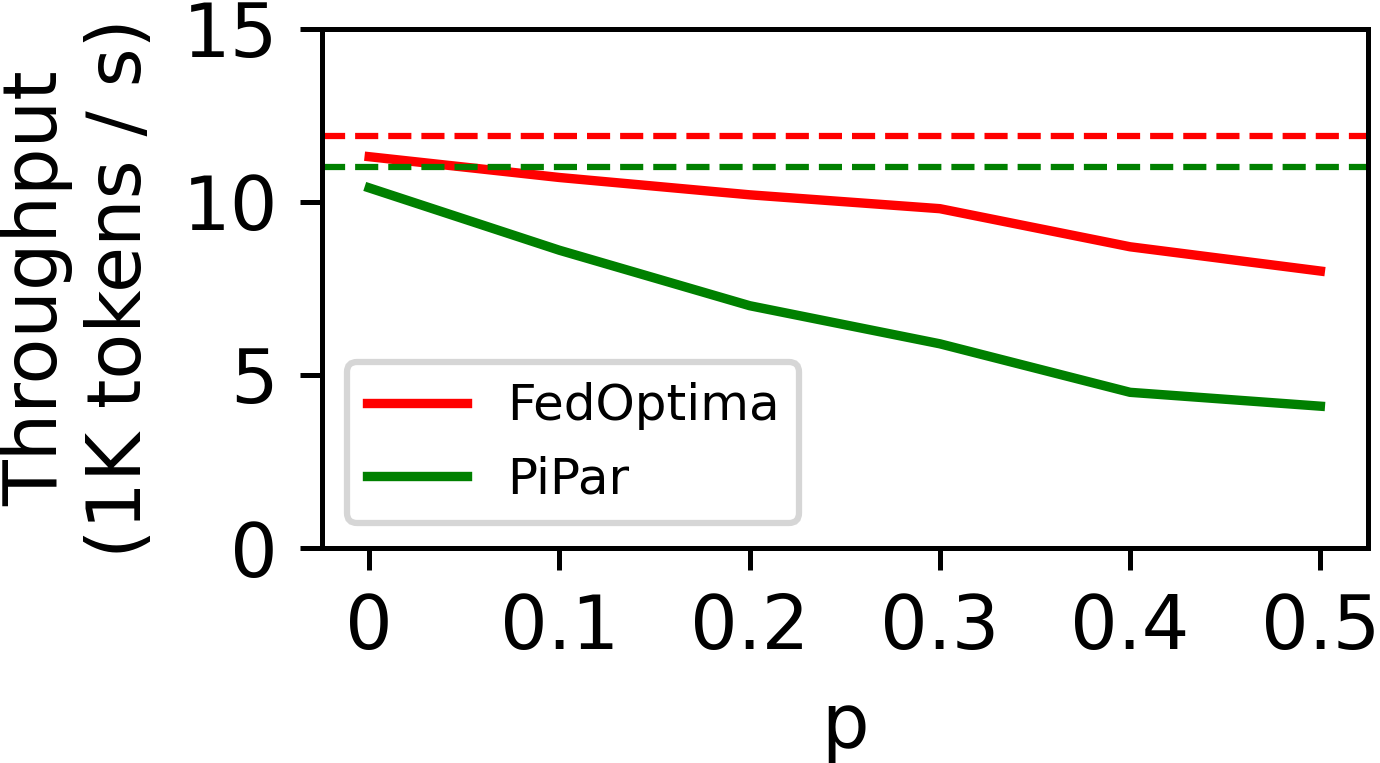}
	       \label{subfig:cpu_nlp_throughput_unstable}
	    }
        \subfigure[Testbed B]{
	    \includegraphics[width=0.35\textwidth]{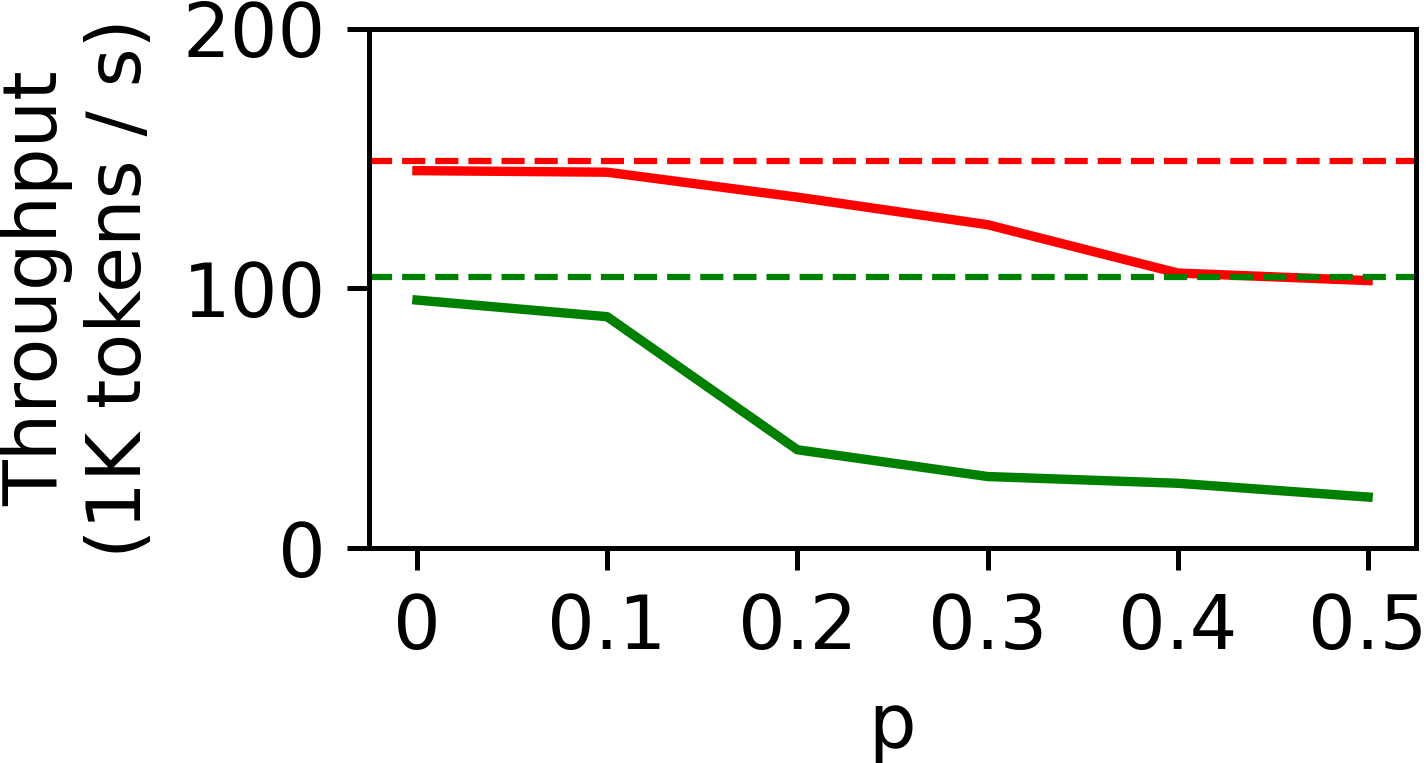}
	       \label{subfig:gpu_nlp_throughput_unstable}
	    }
	\caption{\di{Throughput (higher is better) in an unstable network environment with different probabilities of device leaving ($p$) for sentiment analysis.}}
        \label{fig:throughput-nlp-unstable}
\end{figure*}

\di{Figure~\ref{fig:throughput-cv-unstable} and Figure~\ref{fig:throughput-nlp-unstable} shows the change to throughput with different $p$ for image classification and sentiment analysis, respectively. The horizontal dashed line indicates throughput in the stable setting. When no device leaves ($p=0$), \MethodName\ achieves higher throughput than the baselines on both testbeds and tasks. For image classification (Figure~\ref{fig:throughput-cv-unstable}), throughput decreases as $p$ increases, but \MethodName\ consistently has higher throughput than FedAsync; the gap is wider on the GPU-accelerated Testbed~B. PiPar's throughput drops more sharply as $p$ increases, indicating higher sensitivity to device dropouts: in PiPar, a leaving device blocks training because aggregation requires local models from all devices, whereas in \MethodName\ leaving devices do not affect training on other devices. For sentiment analysis (Figure~\ref{fig:throughput-nlp-unstable}), when $p=0.5$, \MethodName\ achieves retention ratios of $R(0.5)=0.67$ on Testbed~A and $R(0.5)=0.69$ on Testbed~B, which are higher than those of FedAsync ($R(0.5)=0.37$ on Testbed~A and $R(0.5)=0.19$ on Testbed~B).}

\di{Overall, these results indicate that the throughput of \MethodName\ degrades more gracefully when bandwidth varies and devices dropout compared to the throughput achieved in the stable environment (i.e., higher retention ratio). Thus, \MethodName\ demonstrates stronger throughput resilience in unstable environments.}

\begin{boxH}
    \di{\textbf{Observation 4:} \MethodName\ achieves the highest throughput in unstable environments compared to the representative baselines and is less affected by network bandwidth variations and device dropouts.}
\end{boxH}

\di{

\subsection{Ablation study}
\label{subsec:ablation}

We validate the auxiliary network design and task scheduling within \MethodName\ in this section.

\subsubsection{Auxiliary network design}
We consider auxiliary-network variants while keeping the rest of the pipeline unchanged (asynchronous training/aggregation and the counter-based scheduler).

We compare four variants for image classification on both testbeds (Figure~\ref{fig:train-time-cv-an}): (i) \emph{Default (\MethodName)}: one auxiliary layer (same type as the last server-side layer) plus a dense classifier; (ii) \emph{No Aux Network}: devices receive gradients from the server directly (similar to SplitFed); (iii) \emph{Only Aux Classifier}: the auxiliary network contains only a dense classifier; and (iv) \emph{Deep Aux Network}: two auxiliary layers plus a dense classifier.

On Testbed~A (Figure~\ref{subfig:cpu_cv_acc_an}), the default design achieves the best trade-off between convergence speed and final accuracy, reaching nearly 82\% and remaining consistently higher than the variants in the later training stage. \emph{Only Aux Classifier} and \emph{No Aux Network} improve quickly at the beginning but saturate earlier at lower accuracy, showing that removing auxiliary representation learning negatively impacts the final model quality. \emph{Deep Aux Network} can achieve close to the default final accuracy, but converges more slowly and shows larger accuracy fluctuations, indicating that an over-parameterized auxiliary network introduces additional optimization overhead.

On Testbed~B (Figure~\ref{subfig:gpu_cv_acc_an}), where more computation is carried out on the server, the benefit of the default auxiliary design is more pronounced: \MethodName\ reaches the highest final accuracy (~42\%), while the other variants do not achieve higher than 39\%. This confirms that the auxiliary network provides a measurable accuracy gain in both testbeds and is important in the more heterogeneous setting.

\begin{figure*}[tp]
    \centering
	\subfigure[Testbed A]{
	    \includegraphics[width=0.35\textwidth]{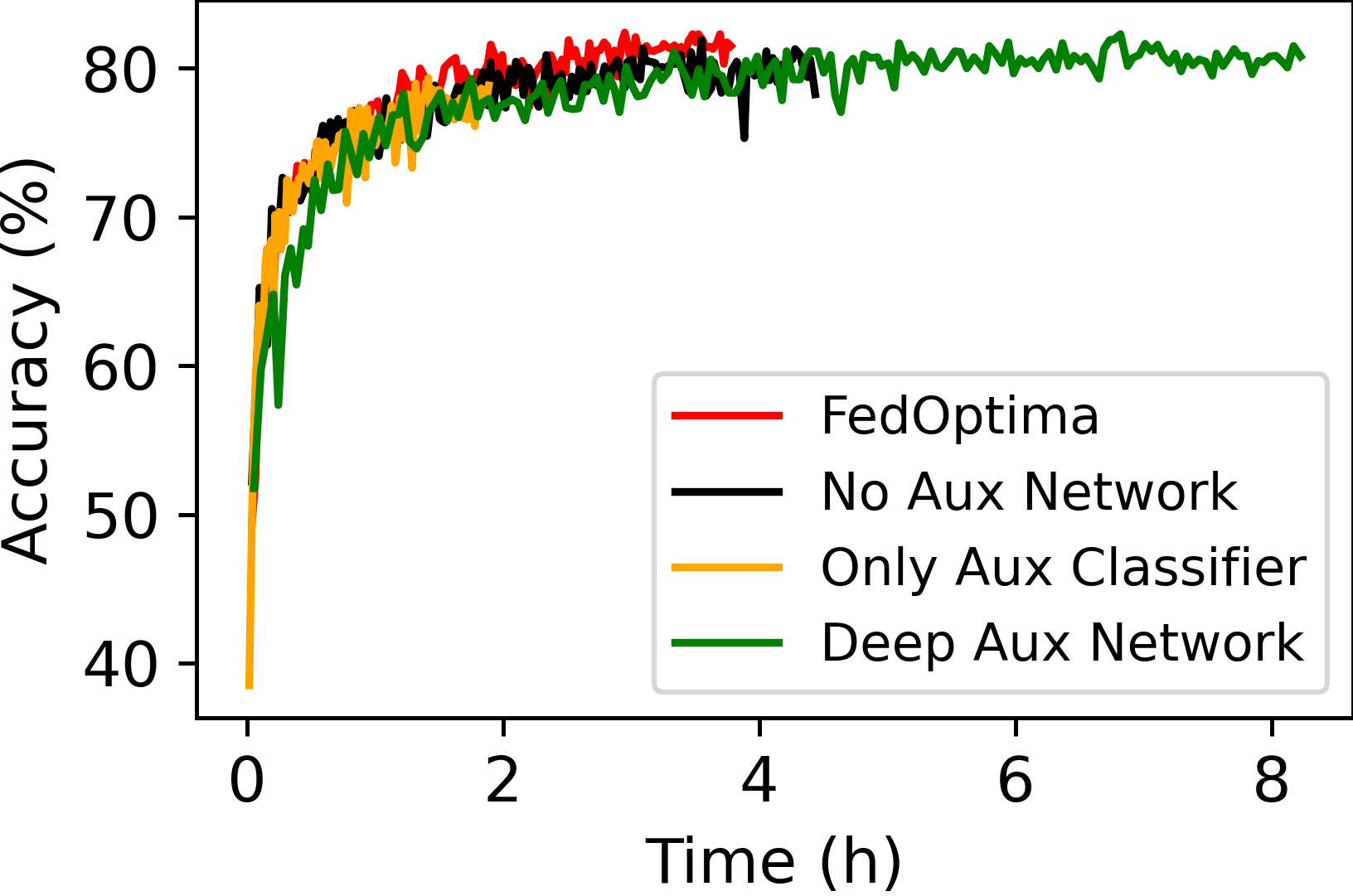}
	    \label{subfig:cpu_cv_acc_an}
	}
	\subfigure[Testbed B]{
		\includegraphics[width=0.35\textwidth]{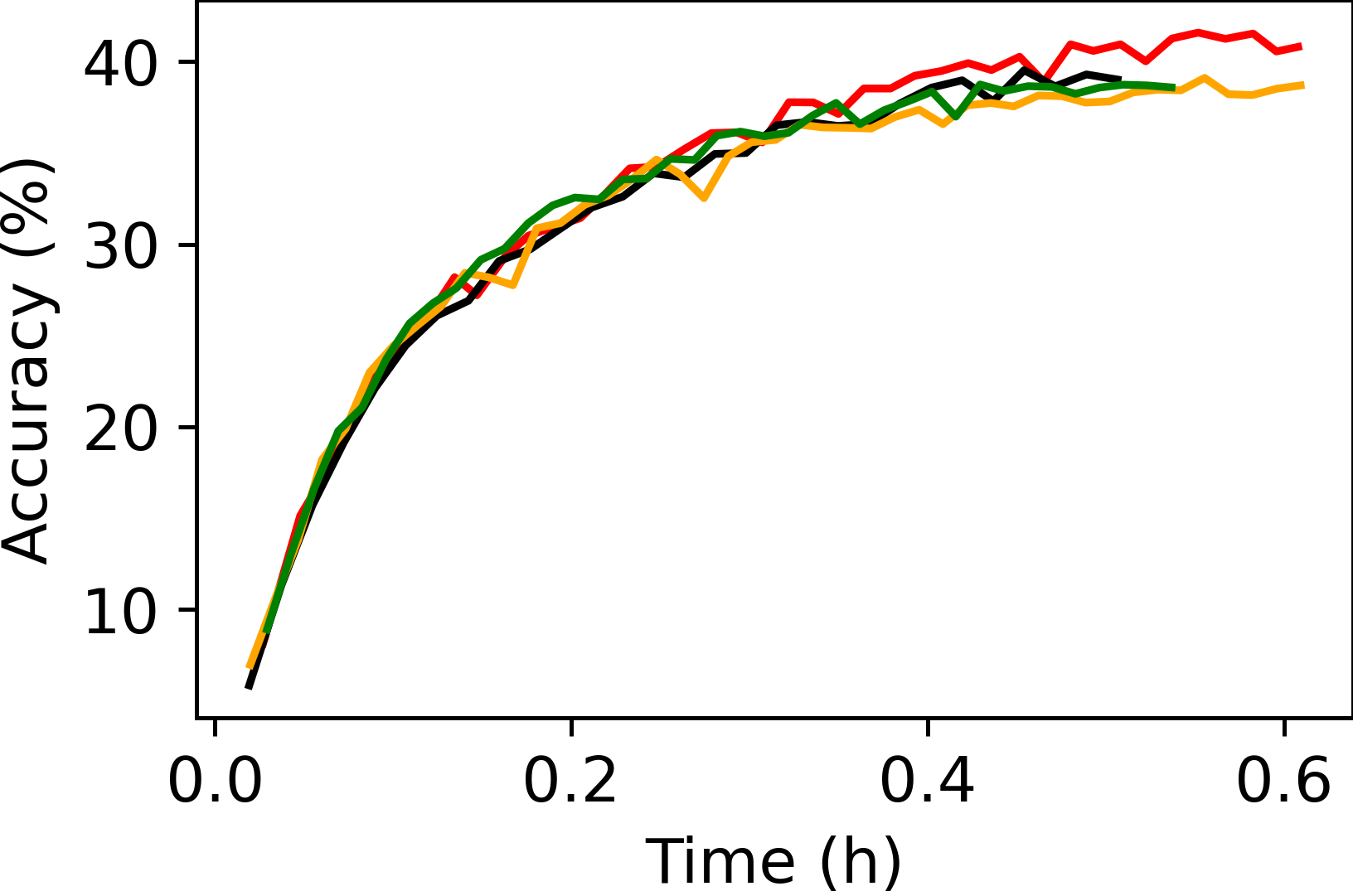}
	    \label{subfig:gpu_cv_acc_an}
	}
	\caption{\di{Accuracy (higher is better) versus training time (lower is better) for image classification with different auxiliary network design.}}
	\label{fig:train-time-cv-an}
\end{figure*}

\begin{boxH}
    \textbf{Observation 5:} For asynchronous training, the default auxiliary-network design of \MethodName\ provides the best convergence-accuracy trade-off. Removing or over-simplifying the auxiliary network lowers final accuracy.
\end{boxH}


\subsubsection{Task scheduling mechanisms}
We further isolate the impact of the task scheduler. We compare the default \emph{counter-based} scheduler in \MethodName\ with a \emph{FIFO} scheduler for image classification on both testbeds (Figure~\ref{fig:train-time-cv-ts}). All other components, including asynchronous aggregation and the default auxiliary network, are kept identical.

On Testbed~A (Figure~\ref{subfig:cpu_cv_acc_ts}), both schedulers show similar early-stage behavior, but the counter-based scheduler achieves slightly better late-stage accuracy (~1\% higher at convergence). On Testbed~B (Figure~\ref{subfig:gpu_cv_acc_ts}) where there are more heterogeneous devices, the gap becomes larger: the counter-based scheduler consistently stays above FIFO and reaches a clearly higher final accuracy. FIFO tends to over-consume activations from fast devices, which increases bias and staleness under heterogeneity; counter-based scheduler mitigates this by keeping device contributions more balanced.

\begin{figure*}[tp]
    \centering
	\subfigure[Testbed A]{
		\includegraphics[width=0.35\textwidth]{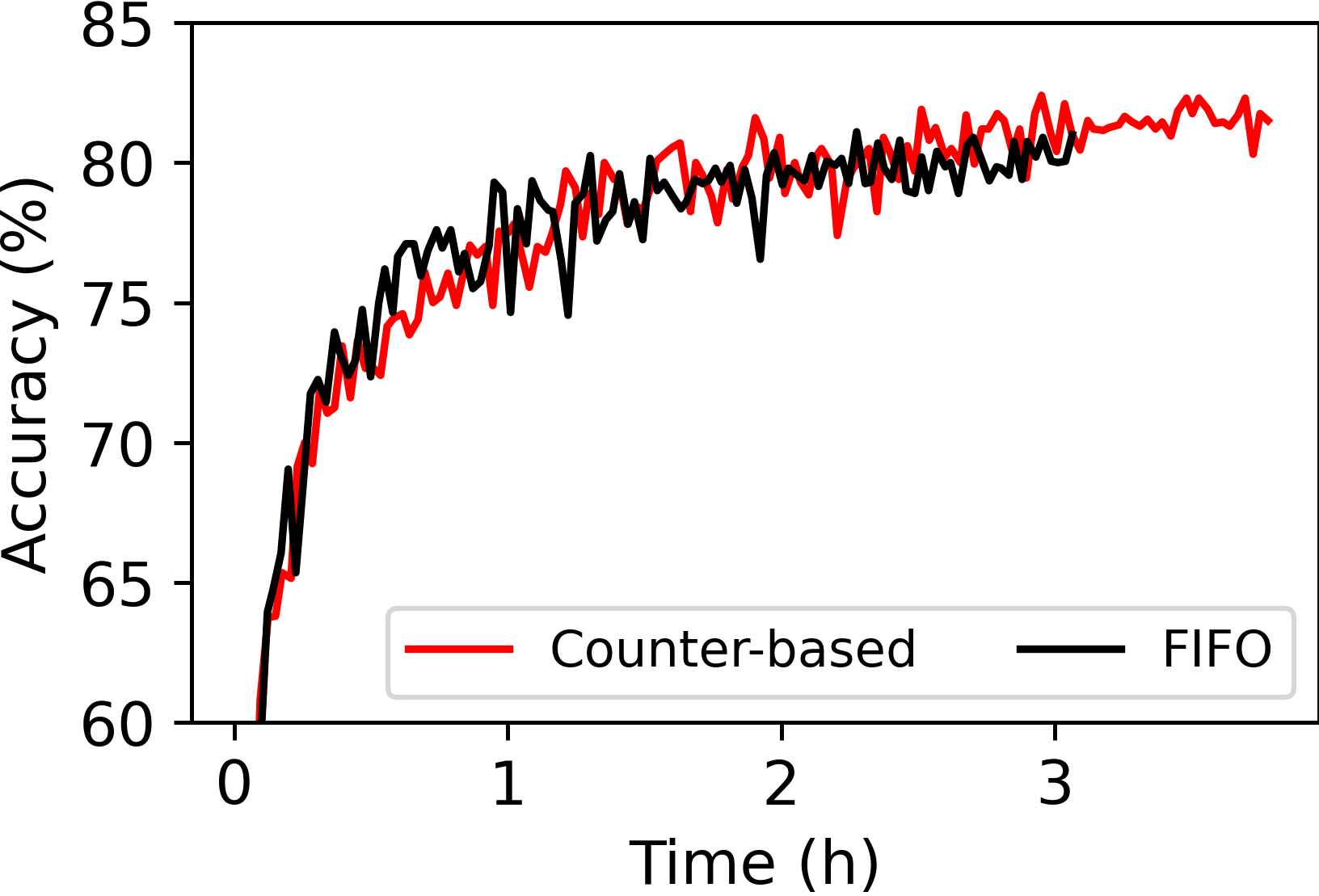}
	    \label{subfig:cpu_cv_acc_ts}
	}
	\subfigure[Testbed B]{
	    \includegraphics[width=0.35\textwidth]{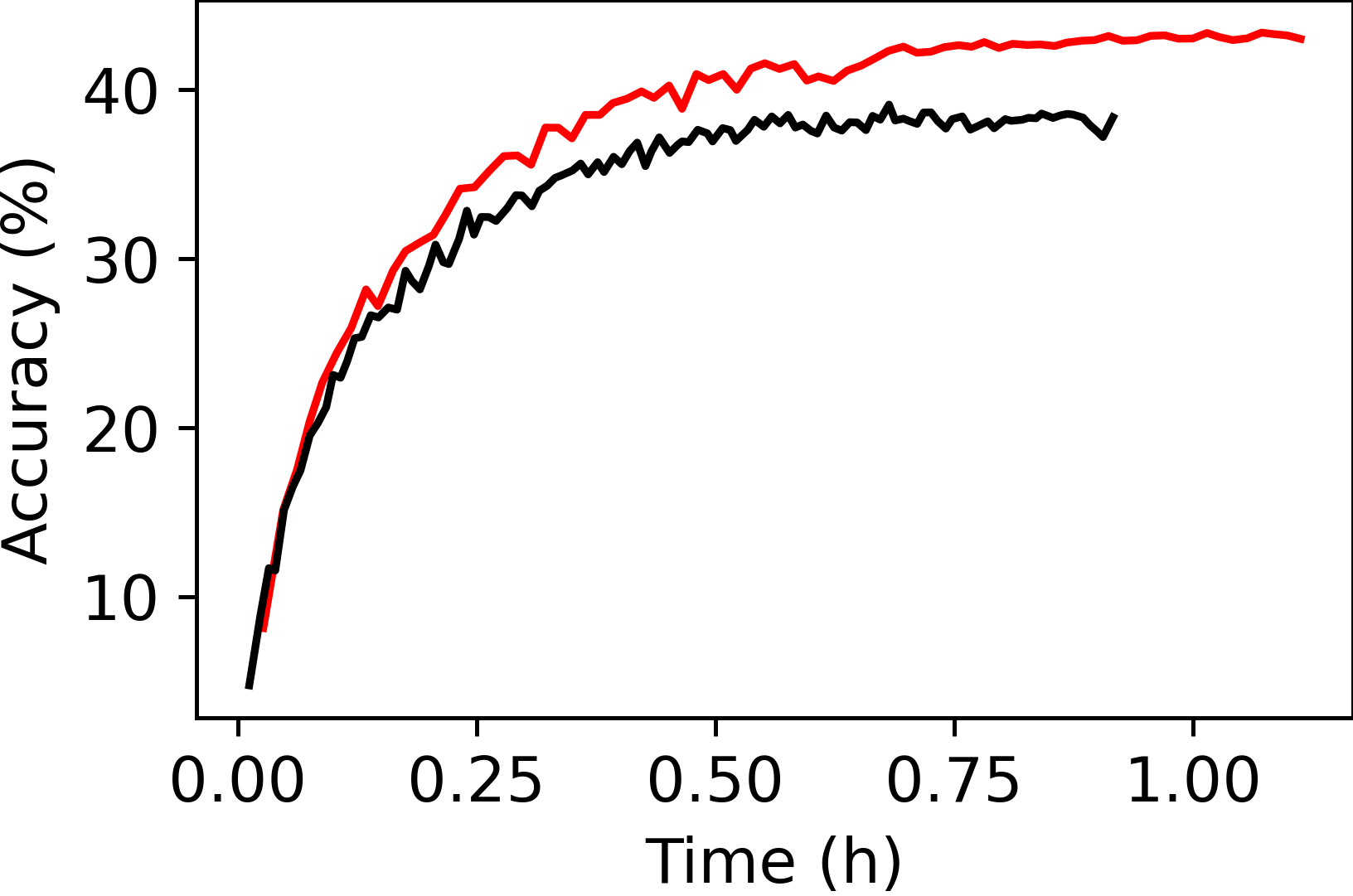}
	    \label{subfig:gpu_cv_acc_ts}
	}
		\caption{\di{Accuracy (higher is better) versus training time (lower is better) for image classification with different task scheduling mechanisms.}}
        \label{fig:train-time-cv-ts}
\end{figure*}

\begin{boxH}
    \textbf{Observation 6:} The counter-based scheduler of \MethodName\ consistently improves accuracy over the FIFO scheduler, with larger gains on the more heterogeneous testbed.
\end{boxH}

}

\section{Related work}
\label{sec:rw}
In classic FL~\cite{fedavg}, the entire neural network is trained on a device, which results in server idle time. Offloading-based FL methods improve server utilization. Split learning (SL)~\cite{sl,slh} was the first work to offload a part of the neural network to the server. The first device trains the initial layers and sends activations to the server that trains the subsequent layers. Then the device-side network is sent to the next device. Training performance is low since devices work in a round-robin.

SplitFed~\cite{splitfed} combines SL and FL to split the neural network across the server and devices and train all devices in parallel. Although it improves server utilization, the dependency between server and device training results in idle time. PiPar~\cite{pipar} further reduces server idle time by using pipeline parallelism to optimize computational efficiency of FL. CSE-FSL~\cite{cse-fsl} and SplitGP~\cite{splitgp} use the auxiliary network and local loss on devices to mitigate task dependency. Since devices compute gradients, the server does not send gradients to devices, which reduces the communication overhead. FedGKT~\cite{fedgkt} transfers logit tensors between server and devices to guide training on both sides. An incentive mechanism is proposed to encourage client participation and enhance model accuracy.~\cite{10631047}. More recent works such as Split-Mix~\cite{splitmix2022} introduce an efficient framework that supports on-demand and in-situ model customization through feature and class partitioning. RoS-FL~\cite{roscfl2022} enhances robustness of split learning in medical image analysis by dynamically correcting the importance weights of device models. However, these methods do not address device stragglers.

To address the straggler problem, FedAsync~\cite{fedasync} employs asynchronous aggregation in FL. Aggregation occurs when any device completes a training round, and a device obtains the global model without waiting for other devices. However, asynchronous aggregation introduces staleness that impacts model accuracy. FedBuff~\cite{fedbuff} balances stragglers and staleness using a buffering mechanism. The server aggregates updates from a subset of devices rather than all devices. Nevertheless, FedBuff also faces model staleness. Libra~\cite{10631034} improves client scheduling in asynchronous FL to mitigate staleness. FedStaleWeight~\cite{fedstaleweight2024} improves fairness and convergence in asynchronous FL by reweighting stale updates based on their staleness degree. SC-AFL~\cite{scaflafl2024} introduces a staleness threshold mechanism to balance accuracy and communication in asynchronous aggregation. However, the above methods do not improve server utilization. \di{More broadly, resilience and recovery in complex traffic networks has also been studied using trend forecasting~\cite{trafficresilience2025}; this line of work is complementary but targets a different system domain than FL.}

\di{Existing auxiliary-network designs in split learning and offloading-based FL are often motivated by mitigating task dependency or enabling device-specific objectives (e.g., personalization). In contrast, \MethodName\ uses the auxiliary network to remove the server-to-device gradient path, which makes gradient-free offloading compatible with asynchronous device-side aggregation. As a result, \MethodName\ simultaneously addresses task dependency and stragglers under heterogeneous devices and non-IID data settings.}


\section{Conclusion}
\label{sec:conclusion}
Classic federated learning under-utilizes computational resources, resulting in server and device idle times and low training efficiency. Two types of idle time stem from server-device task dependency (Type~I) and device stragglers (Type~II). \MethodName, a novel resource-optimized federated learning system is developed to simultaneously minimize both types of idle time; existing research can reduce either but not both at the same time.
\MethodName\ redesigns the distributed training of a DNN. Devices train the initial layers of the DNN, and the rest of the layers is offloaded to the server. Using an auxiliary network and asynchronous aggregation, training on a device is decoupled from the server and other devices, which eliminates Type~II idle time. The server undertakes centralized training on the activations received from devices for reducing Type~I idle time. A task scheduler is used to balance the contribution of heterogeneous devices. The server memory is efficiently managed by controlling the activation flow to enhance the scalability of the system.
Compared to the best results of four state-of-the-art offloading-based and asynchronous FL baselines, \MethodName\ achieves higher or comparable accuracy, accelerates training by 1.9$\times$ to 21.8$\times$, reduces server and device idle time by up to 93.9\% and 81.8\%, respectively, and increases throughput by 1.1$\times$ to 2.0$\times$.


\section*{Acknowledgments}
This research is funded by Rakuten Mobile, Inc., Japan, and supported by funding from UK Research and Innovation grant EP/Y028813/1.

\bibliographystyle{elsarticle-num}
\bibliography{ref.bib}

@article{trafficresilience2025,
  author  = {Hong, Sheng and Yue, Tianyu and You, Yang and Lv, Zhengnan and Tang, Xu and Hu, Jing and Yin, Hongwei},
  title   = {{A Resilience Recovery Method for Complex Traffic Network Security Based on Trend Forecasting}},
  journal = {International Journal of Intelligent Systems},
  year    = {2025},
  volume  = {2025},
  number  = {1},
  pages   = {3715086},
  doi     = {10.1155/int/3715086},
}

@article{splitfed,
  author    = {Chandra Thapa and
               Mahawaga Arachchige Pathum Chamikara and
               Seyit Camtepe},
  title     = {{SplitFed: When Federated Learning Meets Split Learning}},
  journal   = {AAAI Conference on Artificial Intelligence},
  volume    = {36(8)},
  year      = {2022},
  pages      = {8485-8493},
}

@inproceedings{fedasync,
    author= {Cong Xie and Oluwasanmi Koyejo and Indranil Gupta},
    title= {{Asynchronous Federated Optimization}},
    booktitle = {12th Annual Workshop on Optimization for Machine Learning},
    year = {2023},
}

@article{pipar,
    title = {{PiPar: Pipeline parallelism for collaborative machine learning}},
    journal = {Journal of Parallel and Distributed Computing},
    volume = {193},
    pages = {104947},
    year = {2024},
    author = {Zihan Zhang and Philip Rodgers and Peter Kilpatrick and Ivor Spence and Blesson Varghese},
}

@InProceedings{fedbuff,
  title = 	 {{Federated Learning with Buffered Asynchronous Aggregation}},
  author =       {Nguyen, John and Malik, Kshitiz and Zhan, Hongyuan and Yousefpour, Ashkan and Rabbat, Mike and Malek, Mani and Huba, Dzmitry},
  booktitle = 	 {Proceedings of The 25th International Conference on Artificial Intelligence and Statistics},
  pages = 	 {3581--3607},
  year = 	 {2022},
  volume = 	 {151},
}

@inproceedings{decoupled,
author = {Belilovsky, Eugene and Eickenberg, Michael and Oyallon, Edouard},
title = {{Decoupled greedy learning of CNNs}},
year = {2020},
booktitle = {Proceedings of the 37th International Conference on Machine Learning},
articleno = {69},
numpages = {10},
}

@InProceedings{fedavg,
  title = 	 {{Communication-Efficient Learning of Deep Networks from Decentralized Data}},
  author = 	 {McMahan, Brendan and Moore, Eider and Ramage, Daniel and Hampson, Seth and Arcas, Blaise Aguera y},
  booktitle = 	 {20th International Conference on Artificial Intelligence and Statistics},
  volume = 	 {54},
  pages = 	 {1273--1282},
  year = 	 {2017},
}

@article{DBLP:journals/corr/KonecnyMR15,
  author    = {Jakub Kone{\v{c}}n{\'y} and
               Brendan McMahan and
               Daniel Ramage},
  title     = {{Federated Optimization: Distributed Optimization Beyond the Datacenter}},
  journal   = {8th NIPS Workshop on Optimization for Machine Learning},
  year      = {2015},
}

@article{DBLP:journals/corr/KonecnyMRR16,
  author    = {Jakub Kone{\v{c}}n{\'y} and
               H. Brendan McMahan and
               Daniel Ramage and
               Peter Richt{\'{a}}rik},
  title     = {{Federated Optimization: Distributed Machine Learning for On-Device Intelligence}},
  journal   = {CoRR},
  volume    = {abs/1610.02527},
  year      = {2016},
}

@article{sl,
  author    = {Otkrist Gupta and
               Ramesh Raskar},
  title     = {{Distributed Learning of Deep Neural Network over Multiple Agents}},
  journal   = {Journal of Network and Computer Applications},
  volume = {116},
  pages = {1-8},
  year  = {2018},
}

@inproceedings{slh,
    author = {Praneeth Vepakomma and
               Otkrist Gupta and
               Tristan Swedish and
               Ramesh Raskar},
    title = {{Split Learning For Health: Distributed Deep Learning without Sharing Raw Patient Data}},
    booktitle = {ICLR Workshop on AI for Social Good},
    year = {2019},
}

@INPROCEEDINGS{splitgp,
  author={Han, Dong-Jun and Kim, Do-Yeon and Choi, Minseok and Brinton, Christopher G. and Moon, Jaekyun},
  booktitle={IEEE Conference on Computer Communications}, 
  title={{SplitGP: Achieving Both Generalization and Personalization in Federated Learning}}, 
  year={2023},
  volume={},
  number={},
  pages={1-10},
}

@article{cse-fsl,
  title={{Communication and Storage Efficient Federated Split Learning}},
  author={Yujia Mu and Cong Shen},
  journal={IEEE International Conference on Communications},
  year={2023},
  pages={2976-2981},
}

@inproceedings{10.1145/3485730.3493444,
author = {Zhang, Tuo and He, Chaoyang and Ma, Tianhao and Gao, Lei and Ma, Mark and Avestimehr, Salman},
title = {{Federated Learning for Internet of Things}},
year = {2021},
booktitle = {Proceedings of the 19th ACM Conference on Embedded Networked Sensor Systems},
pages = {413–419},
}

@ARTICLE{9762360,
  author={Jiang, Yuang and Wang, Shiqiang and Valls, Víctor and Ko, Bong Jun and Lee, Wei-Han and Leung, Kin K. and Tassiulas, Leandros},
  journal={IEEE Transactions on Neural Networks and Learning Systems}, 
  title={{Model Pruning Enables Efficient Federated Learning on Edge Devices}}, 
  year={2023},
  volume={34},
  number={12},
  pages={10374-10386}
}

@ARTICLE{fedadapt,
  author={Wu, Di and Ullah, Rehmat and Harvey, Paul and Kilpatrick, Peter and Spence, Ivor and Varghese, Blesson},
  journal={IEEE Internet of Things Journal}, 
  title={{FedAdapt: Adaptive Offloading for IoT Devices in Federated Learning}}, 
  year={2022},
  volume={9},
  number={21},
  pages={20889-20901},
}

@article{vgg,
  author = {Simonyan, Karen and Zisserman, Andrew},
  journal = {3rd International Conference on Learning Representations},
  title = {{Very Deep Convolutional Networks for Large-scale Image Recognition}},
  year = {2015},
  pages = {1–14}
}

@article{mobilenet,
  author    = {Andrew Howard and
               Mark Sandler and
               Grace Chu and
               Liang{-}Chieh Chen and
               Bo Chen and
               Mingxing Tan and
               Weijun Wang and
               Yukun Zhu and
               Ruoming Pang and
               Vijay Vasudevan and
               Quoc V. Le and
               Hartwig Adam},
  title     = {{Searching for MobileNetV3}},
  journal   = {IEEE/CVF International Conference on Computer Vision},
  year      = {2019},
  pages  = {1314-1324}
}

@misc{cifar10,
title= {{CIFAR-10 (Canadian Institute for Advanced Research)}},
author= {Alex Krizhevsky and Vinod Nair and Geoffrey Hinton},
year = {2009}
}

@article{cifar10-2,
  title={{Learning Multiple Layers of Features from Tiny Images}},
  author={Krizhevsky, A. and Hinton, G.},
  journal={Master's thesis, Department of Computer Science, University of Toronto},
  year={2009},
}

@article{tinyimagenet,
  author       = {Olga Russakovsky and
                  Jia Deng and
                  Hao Su and
                  Jonathan Krause and
                  Sanjeev Satheesh and
                  Sean Ma and
                  Zhiheng Huang and
                  Andrej Karpathy and
                  Aditya Khosla and
                  Michael S. Bernstein and
                  Alexander C. Berg and
                  Li Fei{-}Fei},
  title        = {{ImageNet Large Scale Visual Recognition Challenge}},
  journal      = {CoRR},
  volume       = {abs/1409.0575},
  year         = {2014},
}

@inproceedings{fedgkt,
author = {He, Chaoyang and Annavaram, Murali and Avestimehr, Salman},
title = {{Group Knowledge Transfer: Federated Learning of Large CNNs at the Edge}},
year = {2020},
booktitle = {Proceedings of the 34th International Conference on Neural Information Processing Systems},
numpages = {13},
}

@inproceedings{transformer,
 author = {Vaswani, Ashish and Shazeer, Noam and Parmar, Niki and Uszkoreit, Jakob and Jones, Llion and Gomez, Aidan N and Kaiser, \L ukasz and Polosukhin, Illia},
 booktitle = {Advances in Neural Information Processing Systems},
 title = {{Attention is All you Need}},
 volume = {30},
 year = {2017}
}

@InProceedings{imdb,
  author    = {Maas, Andrew L.  and  Daly, Raymond E.  and  Pham, Peter T.  and  Huang, Dan  and  Ng, Andrew Y.  and  Potts, Christopher},
  title     = {{Learning Word Vectors for Sentiment Analysis}},
  booktitle = {Proceedings of the 49th Annual Meeting of the Association for Computational Linguistics: Human Language Technologies},
  month     = {June},
  year      = {2011},
  pages     = {142--150},
}

@inproceedings{sst2,
    title = {{Recursive Deep Models for Semantic Compositionality Over a Sentiment Treebank}},
    author = {Socher, Richard  and
      Perelygin, Alex  and
      Wu, Jean  and
      Chuang, Jason  and
      Manning, Christopher D.  and
      Ng, Andrew  and
      Potts, Christopher},
    booktitle = {Proceedings of the 2013 Conference on Empirical Methods in Natural Language Processing},
    year = {2013},
    pages = {1631--1642},
}

@INPROCEEDINGS{resnet,
  author={He, Kaiming and Zhang, Xiangyu and Ren, Shaoqing and Sun, Jian},
  booktitle={IEEE Conference on Computer Vision and Pattern Recognition}, 
  title={{Deep Residual Learning for Image Recognition}}, 
  year={2016},
  pages={770-778},
}

@inproceedings{
acar2021federated,
title={{Federated Learning Based on Dynamic Regularization}},
author={Durmus Alp Emre Acar and Yue Zhao and Ramon Matas and Matthew Mattina and Paul Whatmough and Venkatesh Saligrama},
booktitle={International Conference on Learning Representations},
year={2021}
}

@inproceedings {oort,
author = {Fan Lai and Xiangfeng Zhu and Harsha V. Madhyastha and Mosharaf Chowdhury},
title = {{Oort: Efficient Federated Learning via Guided Participant Selection}},
booktitle = {15th {USENIX} Symposium on Operating Systems Design and Implementation},
year = {2021},
pages = {19--35},
}

@inproceedings{refl,
author = {Abdelmoniem, Ahmed M. and Sahu, Atal Narayan and Canini, Marco and Fahmy, Suhaib A.},
title = {{REFL: Resource-Efficient Federated Learning}},
year = {2023},
booktitle = {Proceedings of the Eighteenth European Conference on Computer Systems},
pages = {215–232},
numpages = {18},
}

@INPROCEEDINGS{10631047,
  author={Han, Pengchao and Huang, Chao and Shi, Xingyan and Huang, Jianwei and Liu, Xin},
  booktitle={2024 IEEE 44th International Conference on Distributed Computing Systems}, 
  title={{Incentivizing Participation in SplitFed Learning: Convergence Analysis and Model Versioning}}, 
  year={2024},
  pages={846-856}}

@INPROCEEDINGS {10631034,
author = {Wang, Chun and Huang, Huawei and Li, Ruixin and Liu, Jialiang and Cai, Ting and Zheng, Zibin},
booktitle = {2024 IEEE 44th International Conference on Distributed Computing Systems},
title = {{Libra: A Fairness-Guaranteed Framework for Semi-Asynchronous Federated Learning}},
year = {2024},
pages = {797-808},
}

@inproceedings{dp,
author = {Abadi, Martin and Chu, Andy and Goodfellow, Ian and McMahan, H. Brendan and Mironov, Ilya and Talwar, Kunal and Zhang, Li},
title = {{Deep Learning with Differential Privacy}},
year = {2016},
booktitle = {Proceedings of the 2016 ACM SIGSAC Conference on Computer and Communications Security},
pages = {308–318},
numpages = {11},
}

@INPROCEEDINGS{pixeldp,
  author={Lecuyer, Mathias and Atlidakis, Vaggelis and Geambasu, Roxana and Hsu, Daniel and Jana, Suman},
  booktitle={2019 IEEE Symposium on Security and Privacy (SP)}, 
  title={{Certified Robustness to Adversarial Examples with Differential Privacy}}, 
  year={2019},
  pages={656-672},
}

@article{doi:10.1137/16M1080173,
    author = {Bottou, L\'{e}on and Curtis, Frank E. and Nocedal, Jorge},
    title = {{Optimization Methods for Large-Scale Machine Learning}},
    journal = {SIAM Review},
    volume = {60},
    number = {2},
    pages = {223-311},
    year = {2018},
}

@inproceedings{NEURIPS2018_a36b598a,
     author = {Huo, Zhouyuan and Gu, Bin and Huang, Heng},
     booktitle = {Advances in Neural Information Processing Systems},
     pages = {},
     publisher = {Curran Associates, Inc.},
     title = {{Training Neural Networks Using Features Replay}},
     volume = {31},
     year = {2018}
}

@inproceedings{splitmix2022,
  title={{Efficient Split-Mix Federated Learning for On-Demand and In-Situ Customization}},
  author={Hong, Junyuan and Wang, Haotao and Wang, Zhangyang and Zhou, Jiayu},
  booktitle={ICLR},
  year={2022}
}

@article{roscfl2022,
  title={{Robust Split Federated Learning for U-shaped Medical Image Networks}},
  author={Yang, Ziyuan and Chen, Yingyu and Huangfu, Huijie and others},
  journal={arXiv preprint arXiv:2212.06378},
  year={2022}
}

@article{fedstaleweight2024,
  title={{FedStaleWeight: Buffered Asynchronous Federated Learning with Fair Aggregation via Staleness Reweighting}},
  author={Ma, Jeffrey and Tu, Alan and Chen, Yiling and Reddi, Vijay Janapa},
  journal={arXiv preprint arXiv:2406.02877},
  year={2024}
}

@ARTICLE{scaflafl2024,
  author={Sun, Sheng and Zhang, Zengqi and Pan, Quyang and Liu, Min and Wang, Yuwei and He, Tianliu and Chen, Yali and Wu, Zhiyuan},
  journal={IEEE Transactions on Mobile Computing}, 
  title={{Staleness-Controlled Asynchronous Federated Learning: Accuracy and Efficiency Tradeoff}}, 
  year={2024},
  volume={23},
  number={12},
  pages={12621-12634},
}

@article{prechelt1998earlystopping,
  title={{Automatic Early Stopping Using Cross Validation: Quantifying the Criteria}},
  author={Prechelt, Lutz},
  journal={Neural Networks},
  year={1998},
  volume={11},
  number={4},
  pages={761-767},
  doi={10.1016/S0893-6080(98)00010-0}
}

\end{document}